\documentclass[journal]{IEEEtran}
\usepackage{graphicx}
\usepackage{subfigure}
\usepackage{cite}
\usepackage{epstopdf}
\usepackage{epsfig}
\graphicspath{ {./eps/} }
\usepackage[ruled,vlined]{algorithm2e}
\usepackage{amsmath}
\usepackage{mathtools}
\usepackage{array} % for tables 
\usepackage{longtable} % for tables 
\usepackage[T1]{fontenc}
\usepackage[blocks]{authblk}
\usepackage{algpseudocode}% http://ctan.org/pkg/algorithmicx
\usepackage{lipsum}% http://ctan.org/pkg/lipsum
\usepackage{verbatim}
\usepackage[utf8]{inputenc}
\usepackage[english]{babel}
\usepackage{color}
\usepackage{enumerate}
\usepackage[export]{adjustbox} % % % % Figure  Left or Right 
\usepackage{pifont}% http://ctan.org/pkg/pifont
\usepackage{float}
\floatstyle{plaintop}
\restylefloat{table}

\ifCLASSINFOpdf
\else
\fi
\begin{document}
\title{Socially-Aware Congestion Control in Ad-Hoc Networks: Current Status and The Way Forward}

\author{Hannan Bin Liaqat, Amjad Ali, Junaid Qadir, Ali Kashif Bashir, Muhammad Bilal*, and Fiaz Majeed  
\thanks{H. B. Liaqat and F. Majeed are with the Information Technology Department, University of Gujrat, Gujrat, Pakistan.
E-mail: hannanliaqat@hotmail.com, fiaz.majeed@uog.edu.pk}
\thanks{A. Ali is with the Department of Computer Science, COMSATS University Islamabad, Lahore Campus, Lahore 54000, Pakistan. E-mail: amjad.ali@cuilahore.edu.pk}
\thanks{J. Qadir is with Information Technology University, Lahore, Pakistan. 
 E-mail: junaid.qadir@itu.edu.pk}
\thanks{A. K. Bashir is with the Department of Science and Technology, University of the Faroe Islands, Faroe Islands. E-mail: alib@setur.fo}
\thanks{M. Bilal is with the Division of Computer and Electronic Systems Engineering, Hankuk University of Foreign Studies, Yongin-si, South Korea. E-mail:m.bilal@ieee.org}
\thanks{Corresponding authors: Muhammad Bilal (E-mail: m.bilal@ieee.org)} 
 }

\maketitle
\begin{abstract}
Ad-hoc social networks (ASNETs) represent a special type of traditional ad-hoc network in which a user's social properties (such as the social connections and communications metadata as well as application data) are leveraged for offering enhanced services in a distributed infrastructureless environments. However, the wireless medium, due to limited bandwidth, can easily suffer from the problem of congestion when social metadata and application data are exchanged among nodes---a problem that is compounded by the fact that some nodes may act selfishly and not share its resources. While a number of congestion control schemes have been proposed for the traditional ad-hoc networks, there has been limited focus on incorporating social awareness into congestion control schemes. We revisit the existing traditional ad-hoc congestion control and data distribution protocols and motivate the need for embedding social awareness into these protocols to improve performance. We report that although some work is available in opportunistic network that uses socially-aware techniques to control the congestion issue, this area is largely unexplored and warrants more research attention. In this regards, we highlight the current research progress and identify multiple future directions of research. \end{abstract}

\begin{IEEEkeywords}
transmission control protocol, ad-hoc social networks, congestion control, acknowledgment, opportunistic networks, delay-tolerant networks, bio-inspired 
\end{IEEEkeywords}

\section{Introduction}
The existing social networking applications are being used for communications among people and provide social connectivity through websites. Using these social sites, a large number of people connect with each other on the daily basis. For example, Facebook is the largest social networking website with a population of more than 2 billion as of 2017. Users can access their accounts through their smartphones, laptops and personal digital assistants (PDAs). These applications ensure connectivity among users on the basis of common interests. Users' mutual interests and experiences can be identified through content that is created by their devices automatically. 
Further, they have social properties solely on the Internet: therefore, the usage of these applications is not possible without Internet connectivity. 
 
\par
\begin{figure}[t]
 \centering  
 \includegraphics[width=.48\textwidth]{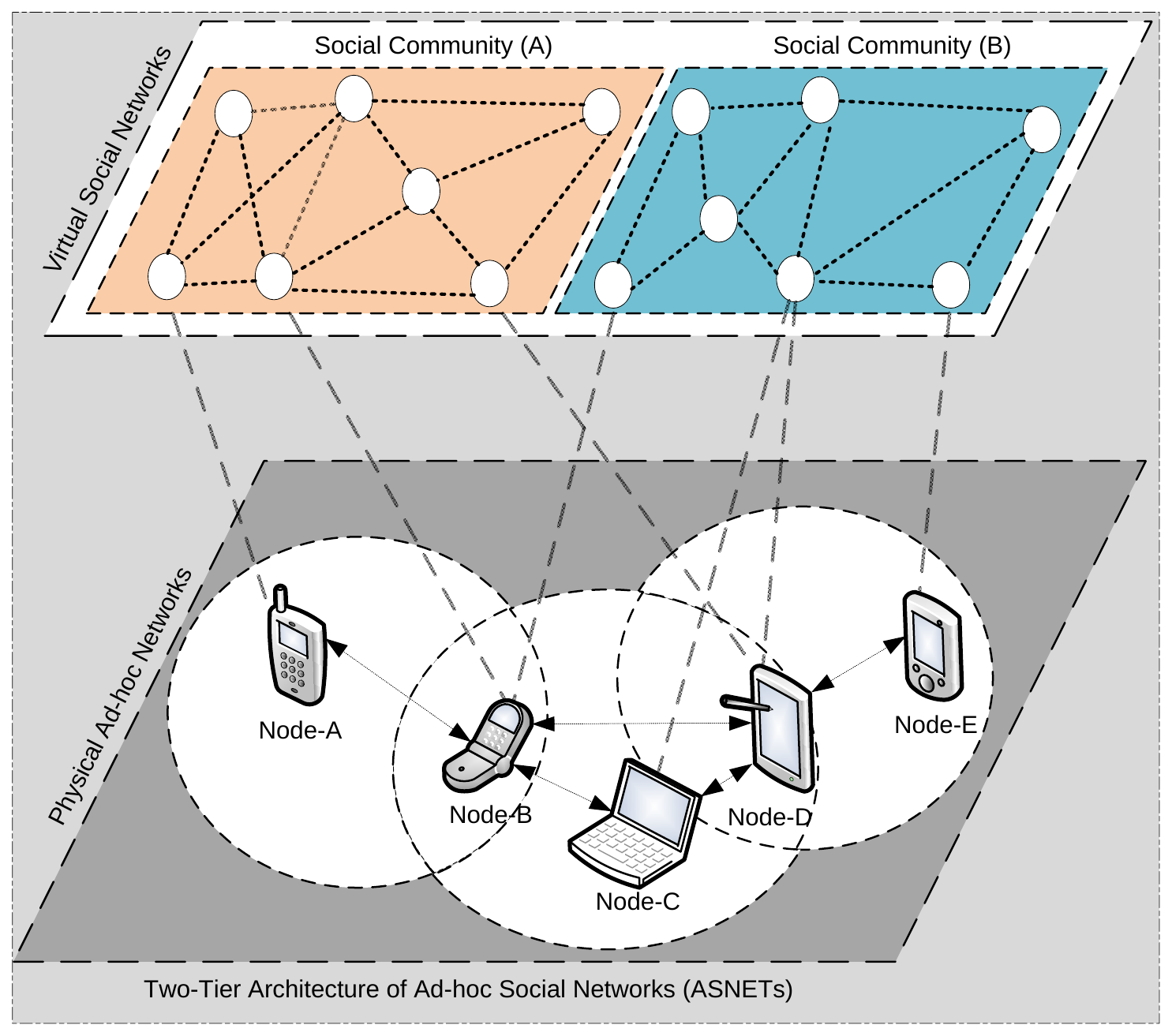}   
 \caption{Convergence of ad-hoc networking with social networks in ASNETs}
  \label{1}
 \end{figure}

 A survey conducted by Wahlstrom et al. in 2014 shows that more than 13 billion wireless fidelity (Wi-Fi) devices will be activated by 2020 \cite{wahlstrom2017smartphone}. Modern mobile devices are equipped with multiple wireless technologies, such as Bluetooth, Wi-Fi, and cellular radio. In designing future networks, the aforementioned technologies help to provide connectivity among users everywhere. For ubiquitous connectivity, the ad-hoc network has tremendous advantages that use the wireless environment for communications. However, the wireless environment has limited resources that require rapid changes in the architectural design of ad-hoc networks. Hence, the concept of ad-hoc social network (ASNET) was introduced in it. The ASNET is a combination of a two-tier architecture based on a physical ad-hoc network and a virtual social network as depicted in Figure~\ref{1}. ASNETs provide advantages in situations when a stable network connection is difficult to create or maintain. Moreover, communications without an ad-hoc network is difficult to achieve in many applications (e.g., in a military environment, a civilian workplace, and in a law enforcement organization). The communication among nodes in ASNETs depends on end-to-end connectivity. However, communication and sharing of resources in ASNETs are quite different from other traditional ad-hoc networks. 

\par In ASNETs, communication depends on the social properties of users ( e.g., social ties, similarities, human mobility patterns, centrality, community, and social graphs), which can be leveraged for the efficient utilization of resources \cite{xia2015socially,katsaros2010social}. Today, people from all walks of life communicate through social networking applications on daily basis for business, education, and entertainment like purposes. Such a growing market of social networks is a major reason behind the motivation to design ASNETs. In the social networking paradigm, due to the larger number of nodes, and furthermore, to reduce congestion in a network, users want to communicate with specific nodes that have a similar social property. Therefore, to find specific users, the social property called \textit{similarity} has a lot of importance. The considerations of \textit{tie-strengths}, \textit{human mobility patterns}, and \textit{social graphs} also provide advantages in finding a specific person or user. In the context of quickly finding and communicating with a previous user in near proximity, \textit{contact history} has the greatest significance. Furthermore, for reducing the complexity of the network, these social properties help in dividing a bigger community into different sub-communities as shown in Figure~\ref{1}. Moreover, if the destination node is not within range, then a social property like \textit{degree centrality} provides a reliable solution for selection of a relay node. Besides, users want to communicate with nodes ubiquitously, even if accessibility to the Internet is not available. Consequently, ad-hoc networks require social properties that provide social communications everywhere and that help in designing ASNETs. Numerous applications involving social properties in an ad-hoc network are described in the literature \cite{hu2015survey} and are available as products, e.g., Jambo Networks, Nokia Sensor, Eiko, and MultiHoc.

\par The ASNET utilize scarce wireless resources for its communications. In ASNETs, two types of data are generally exchanged among nodes: 1) social metadata, and 2) application data. The social metadata is exchanged to search out users with similar interests. Moreover, social metadata updates are used to create connections between similar-interest nodes. However, the application data is exchanged to satisfy the requirements of social applications. In ASNET, every node tries to find those nodes that have similar interests and that share its own profile; otherwise, the node works as a relay node \cite{ciobanu2015interest}. To recognize communicating nodes, the social profile of each node is shared with short handshake messages. If there was a change between the nodes since the last contact, full social profiles are also exchanged. The social profiles of discovered nodes are matched after a completed handshake from the execution of a forwarding algorithm by the social network. A wireless network may easily suffer from the problem of congestion when such extra data is exchanged for communications. 

  \begin{figure*}
  \centering  
  \includegraphics[width=1.05\textwidth]{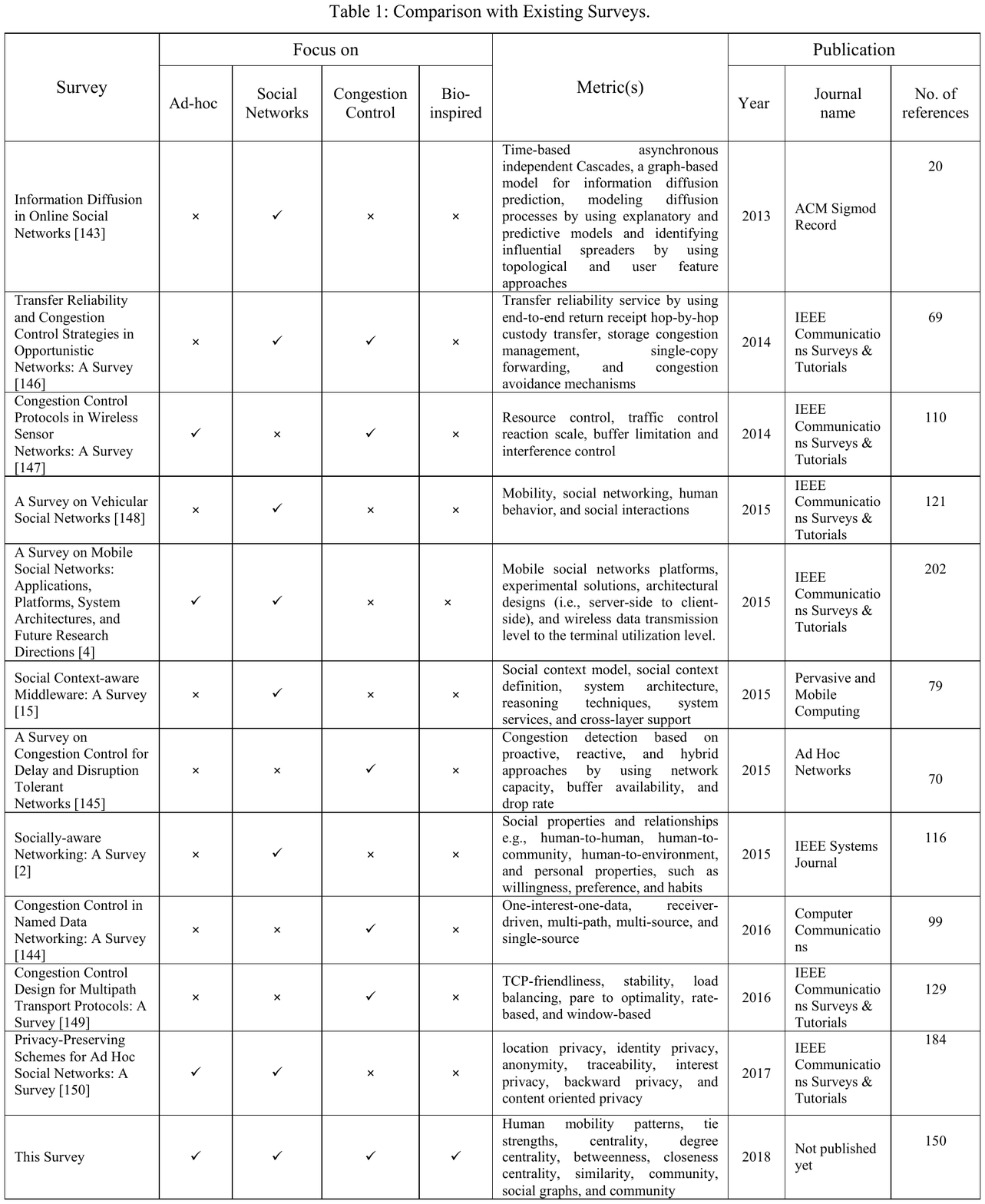}   
  %\caption{Convergence of ad-hoc with social networks.}
 % \vspace{-1.5em}
 % \label{fig:digraph}
  \end{figure*} 
  
\par In ASNETs, due to the simultaneous utilization of several social applications at a single node, and the presence of a large number of users in a single social community, efficient resource sharing among nodes and avoidance of a congestion-related loss becomes vitally important. To use resources efficiently and to provide fairness among data using the social properties of nodes, it is necessary to avoid overwhelming a relay or an intermediate node. In order to reduce losses related to congestion, two kinds of approaches are commonly utilized. One approach is congestion avoidance (CA), and the other is congestion control (CC). The CA activates before congestion occurs because it is a proactive approach. The CC scheme is reactive, and applies when the congestion has occured in the network.

\par The transport layer plays an important role since it controls the data rate before the occurrence of congestion in traditional ad-hoc networks. Furthermore, it needs special care because in ad-hoc networks, connection and disconnection occur frequently, which creates wireless link losses. The solution to congestion-related losses can be provided through many schemes, in which transmission control protocol (TCP), active queue management (AQM) for dropping and scheduling are the most important \cite{eremin2015robust, xu2005tcp,jindal2013efficiency}. Furthermore, a delay in acknowledgments and the partition of bandwidth also provide a huge advantage for solving congestion-related issues. Social network-based nodes use socially-aware properties of users to communicate with each other. Therefore, existing traditional techniques cannot work properly with ASNETs as transferring data in social networks is designed according to the behavior or interests of a user. Furthermore, to resolve congestion issues with respect to social properties, some work has been done for opportunistic networks. However, the working of an opportunistic network does not rely on end-to-end connectivity [145-146]; therefore, the suggested methods do not provide the congestion-related information of the whole network. The reason for the unavailability of nodes for the end-to-end communication can be attributed to longer delays since longer delays impede the identification and the rectification of a node's reliability and congestion issues. Moreover, an opportunistic network only considers the socially-aware node selection and routing techniques to address the congestion issues. However, the ASNETs required end-to-end connectivity between the sender and the destination nodes. Therefore, existing solutions for opportunistic networks are not directly applicable in ASNETs. Congestion solutions for ASNETs should thus, consider the physical aspects (e.g., loss differentiation, shared link, overhead management, congestion window etc.) as well as the social aspects (e.g., tie-strengths, human mobility patterns, social graphs etc.).

\begin{figure*}[t!]
  \centering  
  \includegraphics[width=1\textwidth]{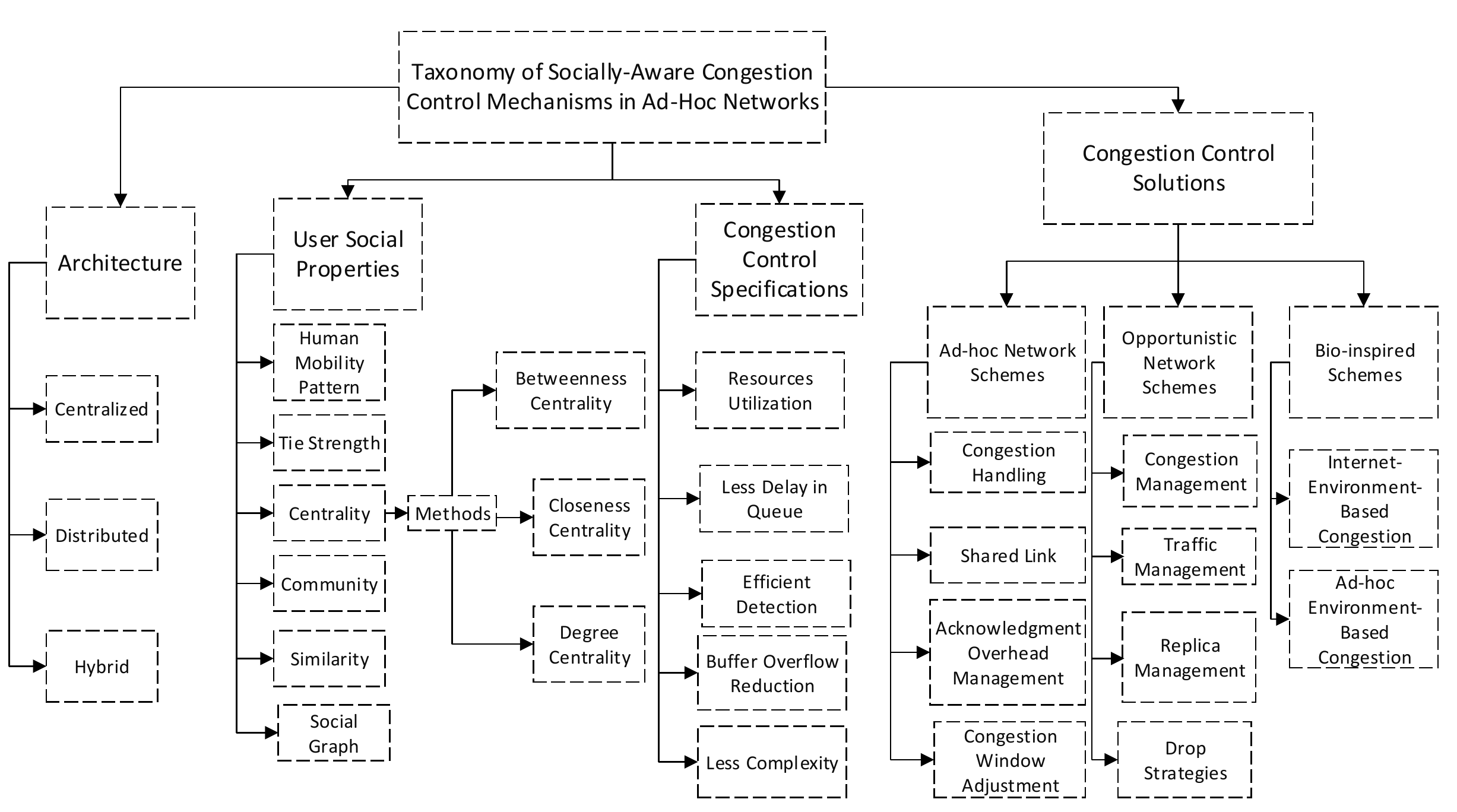}   
  \caption{Taxonomy of socially-aware congestion control mechanisms in Ad-hoc networks }
  \label{2}
  \end{figure*}
 
\par Social properties are not only helpful for efficient resources sharing among nodes, but they also define the mobility patterns efficiently \cite{xia2016overhead,xia2016user,liaqat2015social}. Therefore, for efficient resource utilization in ASNETs, first, the source node calculates available resources by using the traditional methods and then, shares resources among nodes based on social properties. Hence, the aim of this study is to motivate the researchers to design efficient congestion control protocols for ASNETs. Therefore, this paper revisits the congestion control mechanisms that are already addressed in the domain of traditional ad-hoc networks and opportunistic networks but do not resolve the major concerns in ASNETs. In order to understand the concept of socially-aware networks, we first review the detailed architecture of different social networks with their available social properties. After consideration of the users' social properties, we analyze an extensive literature review of existing congestion control schemes that were developed in ad-hoc and opportunistic networks, due to lack of knowledge in ASNETs. However, the pattern of communication is different in ASNETs (due to the involvement of social metadata and application data) from the traditional ad-hoc network. Furthermore, to solve congestion issues in a natural way, some bio-inspired methodologies are reviewed. The consideration of these schemes is useful for designing an efficient and resourceful congestion control mechanism for ASNETs. To classify the existing work in social networks, few developments are discussed with new research challenges. The consideration of congestion-aware protocols is advantageous in resource-scarce environments. However, the existing developments in social networks indicate that little work has focused on resolving congestion issues in ASNETs. Therefore, there is great need to redesign the existing ASNET methods with the combination of congestion approaches. Nevertheless, to the best of our knowledge, none of the existing work has surveyed congestion control methods in heterogeneous networks based on social properties in detail, as shown in Table 1. Moreover, the discussions on future challenges also highlight the problems, when ASNETs works in combination with a congestion control protocol.

\par The remainder of this paper is arranged as follows. In Section II, we present structural designs of social networks. We examine the social properties of users in Section III, and provide specifications of congestion control protocols in Section IV. Section V reviews the existing research into congestion control solutions in different networks and approaches. Some current developments in social networks are described in Section VI. Research challenges and future directions are presented in Section VII, and finally the paper is concluded in Section VIII. A taxonomy of congestion control mechanisms in ad-hoc social networks is illustrated in Figure~\ref{2}.

\section{Structural Designs of Social Networks}
\par Social applications are useful when finding users with similar interests and are helpful in finding their physical locations. To provide connectivity and enhance communications among nodes, socially-aware networks can use three types of architecture: 1) centralized, 2) distributed (or ad-hoc), and 3) hybrid. Discussions of these architectures are useful to identify the advantages and disadvantages in networks. Therefore, in the next subsections, we provide details about how communication is performed among nodes in such settings using figures, examples, and elaboration of current research. 

\subsection{Centralized Social Networks}
\par
\begin{figure}[t!]
  \centering  
  \includegraphics[width=.48\textwidth]{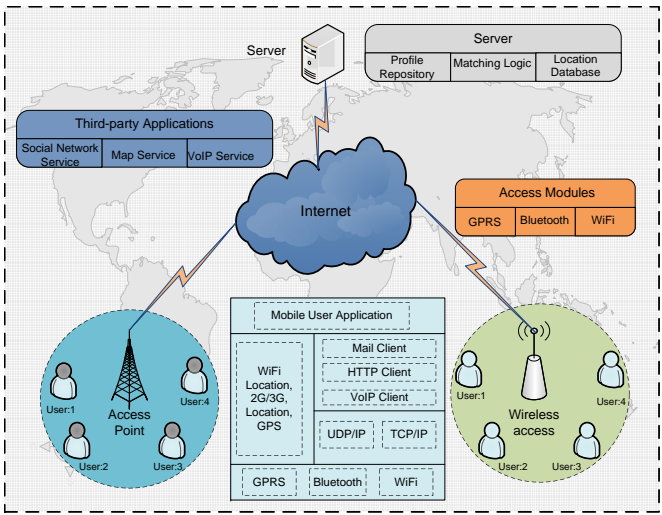}   
  \caption{Centralized architecture of social networks with Internet-based communications}
  %\vspace{-1.5em}
  \label{3}
  \end{figure}

 In centralized social networks, multiple users are connected to one single server; but these nodes cannot communicate directly with each other. The information about all nodes passes through a single centralized server. The relationship among nodes in a centralized architecture is also called one-to-many. For communication to occur in a centralized architecture, these nodes can use wired, wireless, and other similar media interfaces to connect with each other. This centralized server provides information to users through web-based social applications. These social applications collect information from users who are connected to a specific server and, after collecting information, store data in a centralized server. The collected information, such as users' similarities, interests, and positions of the nodes can be used in recommendations for communications and for the transfer of data. Figure~\ref{3} illustrates a detailed communications model of centralized social nodes that communicate through a single server. 

\par To provide communications among nodes, third-party application servers (voice over Internet protocol and the Simple Mail Transport Protocol mail server) can enhance the client/server architecture. According to the concept of the centralized architecture, both advantages and disadvantages coexist in client/server communications. A centralized architecture provides advantages by containing similar information on all connected users. Therefore, when any user changes his/her profile or status, all nodes are easily synchronized, and status is updated in their own profile managers. Nevertheless, a disadvantage of the centralized architecture was elaborated on by Kemp and Reynolds \cite{kemp2009mobile}. Using a hub and spoke model, Kemp and Reynolds noted that all information transferred through a single hub node can create congestion in a network. Furthermore, they also discussed how maximum storage of data at some specific physical location also creates a system bottleneck. 

\par Centralized and web-based architectures work in almost the same manner. In a centralized architecture, mobile nodes are connected and share information through a centralized server. These servers provide connectivity to users by offering some specific mobile browser applications. With the mobile browser, nodes can connect to specific servers, and users can access their own specific accounts, such as Facebook and MySpace. However, to provide ease of access to social applications, a lot of research relating to middleware protocols is ongoing. A middleware deployment of a mobile social computing application called MoBiSoC was presented by Gupta et al. \cite{gupta2009mobisoc}. MoBiSoC uses a people-centric and place-centric approach to divide up data among different servers. To provide access to a centralized application, Ahmed et al. \cite{ahmed2014event} discussed a method called the event-based mobile social network, which combines context-aware mobility with multimedia sharing. Furthermore, detailed analyses of existing middleware-based applications have been presented by Liang et al.  \cite{liang2015social}.

\par Another example of a centralized architecture is the wireless sensor network (WSN) where all sink nodes send data to a single selected node called a cluster head (CH). A sink node acts as a server in a centralized architecture. WSNs can provide personalized and detailed social services using contextual information. Consequently, Kawsar et al. \cite{kawsar2014remote} developed a smartphone application that uses a Hypertext Transfer Protocol (HTTP) based web application programming interface (API) to gather information using the mobile device's sensors. In addition, it also supports transported markup language for push and pull services of published data. However, to design a solution for aspects of social behavior, Jabeur et al. \cite{jabeur2013mobile} discussed a brief literature review of social applications in a centralized architecture. 

\subsection{Distributed (or Ad-hoc) Social Networks}
\par
\begin{figure}[t!]
 \centering  
 \includegraphics[width=.48\textwidth]{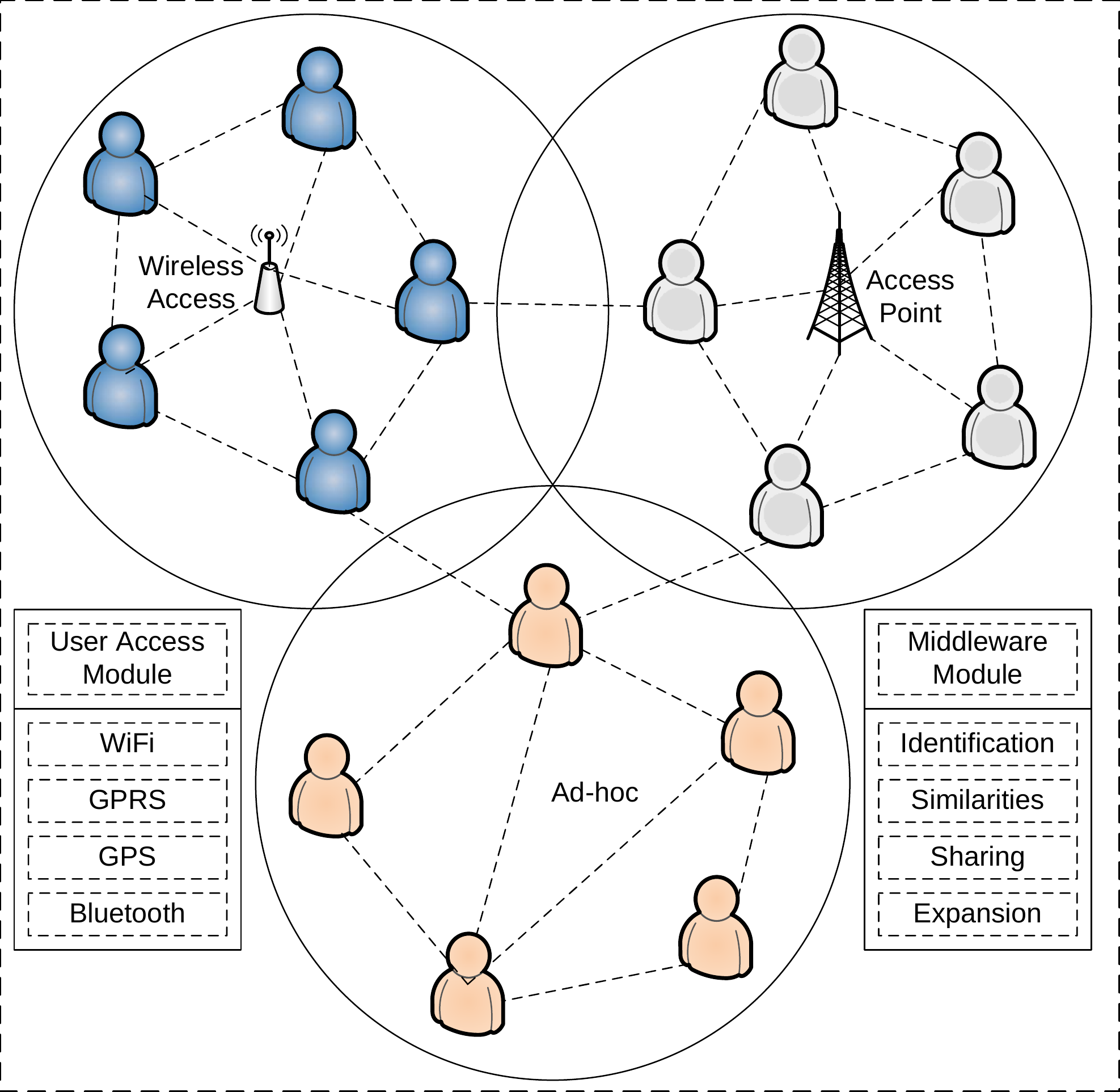}   
 \caption{Distributed architecture of social networks without Internet-based communications}
 \vspace{-1.5em}
 \label{4}
 \end{figure}

 Nodes that work under an architecture of the socially-aware distributed network, communicate without any centralized server. ASNET nodes store social metadata (e.g., social properties) information in their own buffers. Furthermore, these nodes also store social information about their connected neighbor's. Social information is shared among nodes by using the store-and-carry concept. This transfers data to the desired node when a connection is established among nodes. In a distributed architecture, nodes can use both connection-oriented and connection-less environments. This architecture covers two types of network; one is related to the delay tolerant network (DTN) or opportunistic network and the other is the ad-hoc network. In DTNs, nodes have a longer delay between each other for communications; therefore, this network is called a connection-less network. On the other hand, the ad-hoc network depends on nodes connecting with each other before communications start. In this network, the delay among nodes for communications is not longer, so an ad-hoc network is more reliable than an opportunistic network. Both of the above networks are helpful in designing social networks for communications among nodes after using social properties of nodes. Figure~\ref{4} illustrates how distributed socially-aware networks work without the Internet and without any centralized communications.

\par ASNETs provide lower deployment and maintenance costs when a decentralized method is used for communications. Thus, Pietil\"ainen et al. \cite{pietilainen2009mobiclique} developed MobiClique as a middleware for communications. MobiClique is able to provide  content exchange in an ASNET that utilizes user mobility. However, the drawback with MobiClique is that it uses the flooding methodology to disseminate content among users, which makes it less efficient and resource intensive. The division of a decentralized socially-aware network can be defined in one of two methods. The first is deployed by using direct sharing, and the second uses an intermediate node to transfer data. Most researchers are working on the second method, which is based on an intermediate node. This is because the environment of an ad-hoc network is dynamic, and the intermediate node carries data until the destination is found. For that reason, Sarig\"ol et al. \cite{sarigol2009enabling} developed the AdSocial platform for ad-hoc networks that use social applications to transfer data through intermediate nodes. AdSocial was basically developed for small-scale scenarios. It uses already developed routing protocols (any-to-many) to transfer data packets.

\subsection{Hybrid Social Networks} 
\par
\begin{figure}[t!]
 \centering  
 \includegraphics[width=.48\textwidth]{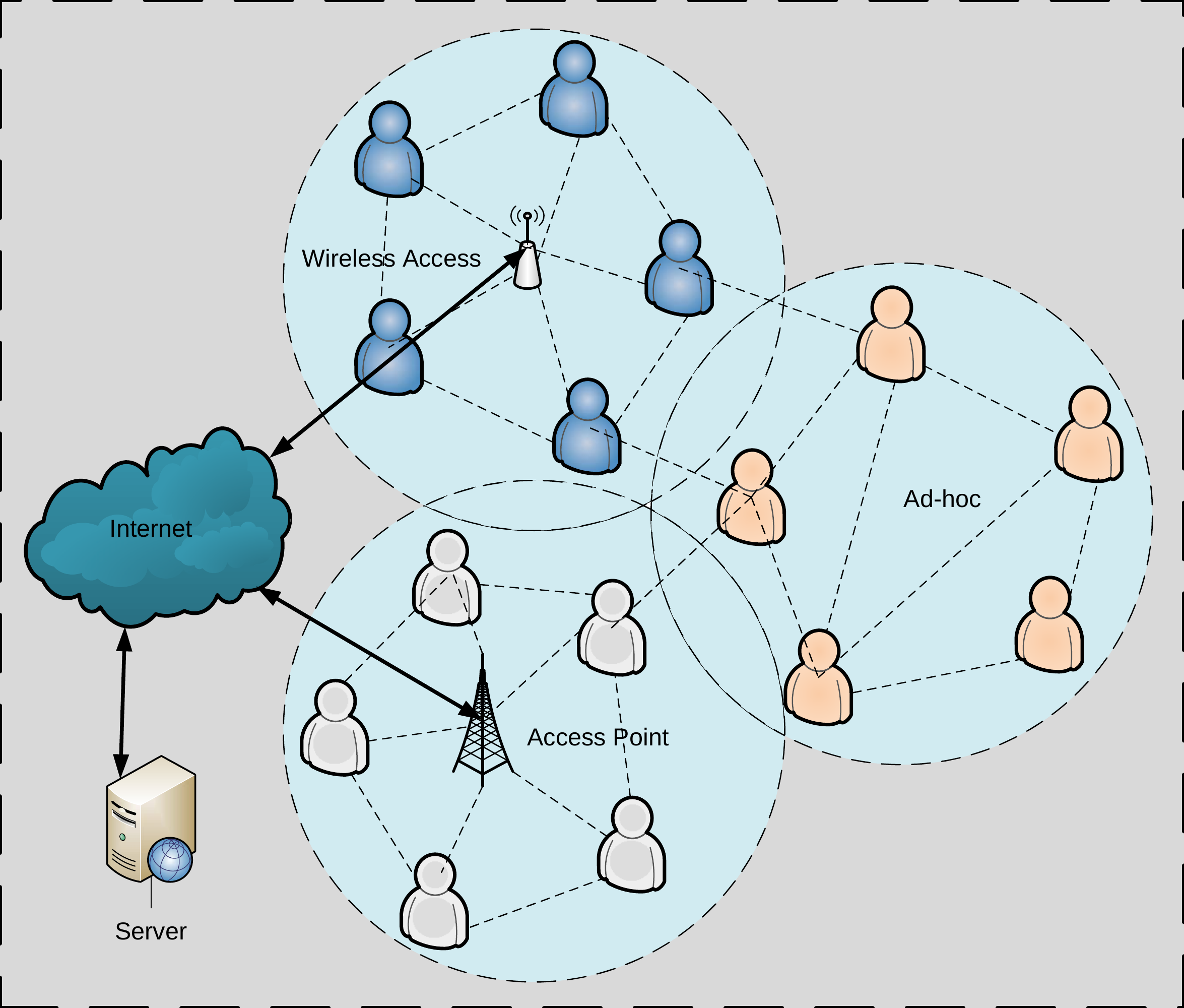}   
 \caption{Hybrid architecture of social networks with/without Internet-based communications}
 \vspace{-1.5em}
 \label{5}
 \end{figure}

 Hybrid networks can access data under both centralized and decentralized architectures. This is helpful in future-generation networks, where a single device has greater accessibility to networks to communicate among nodes. The nodes in a hybrid network use the shortest and lowest cost method for communications. Figure~\ref{5} illustrates the selection of networks for nodes in a hybrid network. The new mobile devices and smartphones have the ability to communicate through both short-range (Bluetooth or wireless) and cellular networks. The architecture of a hybrid network is helpful in providing greater access to, and availability of networks. As a result, Molina et al. \cite{molina2009negotiation} introduced a mobile transient network with access to a multimode device in ad-hoc networks. This multimode device is helpful for communications in both cellular and ad-hoc networks. To disseminate data over similar-interest nodes, an intelligent agent is used to communicate in hybrid networks that use social behavior between users.

\par To provide accessibility to social applications in socially-aware networks, a hybrid platform uses the Jini Surrogate Architecture (JSA) designed by Brooker et al. \cite{brooker2010middleware}. Their experimental results show that a mobile device access both local and cellular technology after the deployment of this middleware. Therefore, a hybrid network architecture provides easy and extensive establishment of a network. Furthermore, to reduce traffic in data from a cellular network, work related to offloading of mobile 3G data was presented \cite{han2012mobile}. 
\section{Users' Social Properties}
\par A social property based communications provides efficient utilization of resources within scarce bandwidth environments. In order to justify social networks, the social properties for communications among nodes were  considered by Milgram \cite{travers1967small}, who introduced the concept of a small world in which users can connect to relationships or small sequences. This work focused primarily on investigating the structural design and dynamics of multiple networks. Moreover, the classification of relationships was studied in social network analysis (SNA) \cite{wasserman1994social}. The usage of SNA in social relationships among users is vital to analyzing complexity levels. To communicate in social networks, networking communities have been attracted by the idea of social networks and schemes for SNA. Hence, in the next sub-sections, we provide in-depth information about social properties that are used for designing algorithms and protocols in socially-aware networks.

\subsection{Human Mobility Pattern}
\par To predict the movement of nodes in the future, human mobility patterns provide an advantage. Researchers have shown that human mobility patterns can be specified through the important properties of spatial and temporal regularity inherent in human mobility. Dependence on spatial regularity normally revolves around a set of frequent locations on a regular time schedule. On the other hand, temporal regularity is that pattern of human mobility that behaves in an iterative manner over a long period of time \cite{hoteit2014estimating,zhao2015n}. The properties mentioned above can be employed to predict the future mobility of users and can play a significant role in selecting nodes as forwarders.

\subsection{Tie Strength}
\par Tie strength, a concept introduced by Granovetter \cite{27} in 1973, is used to characterize two individuals. The popular tie strength indicators (i.e., recency, longevity, multiple social context, frequency, intimacy/closeness, reciprocity, and mutual confiding, (or trust)) were presented by Vastardis et al. \cite{28}. The calculation of the tie strength property is helpful after combining all popular tie strengths and checking for which individual has a strong social relationship with the destination. According to  Granovetter's concept, there may be weak ties that can be useful in forming a link between high-density clusters. 

\subsection{Centrality}
\par
\begin{figure}[t!]
 \centering  
 \includegraphics[width=.48\textwidth]{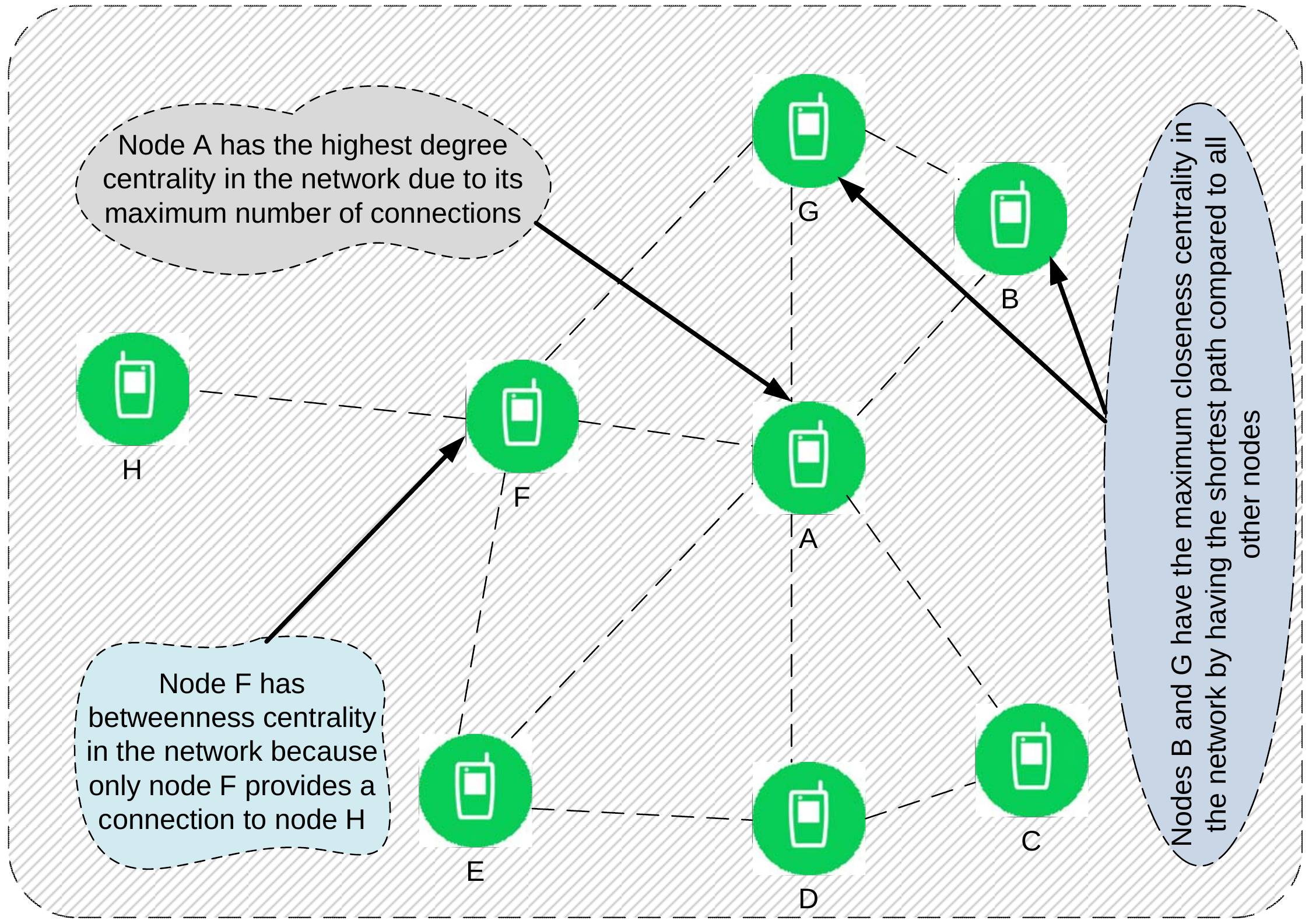}   
 \caption{Communication among nodes using different centrality (such as Degree, Betweenness, Closeness Centrality) models}
 \vspace{-1.5em}
 \label{6}
 \end{figure}

 For socially aware forwarding, the concept of centrality significantly impacts the efficiency of the algorithms. According to Soelistijanto and Hawarth \cite{29}, results showed that relevant centrality and community must be found in order to design data dissemination and socially aware forwarding. In a social graph, the use of centrality helps to explain important nodes. Freeman along with Vazquez-Rodas and Luis introduced multiple methods to measure centralities that are based on degree, betweenness, and closeness \cite{30,31}. A brief introduction to centrality measurement is given in the following sub-sections, and Figure~\ref{6} illustrates the concepts of different centrality models.

\subsubsection{\textbf{Degree Centrality}}
\par This is generally given as the proportion of the maximum number of possible communications to the count of directly connected neighbor nodes, and is used to represent a node's reputation. For an network with $m$ nodes, degree centrality $D(Ca)$ of node $a$ is described as:
\begin{equation}
D(C_a)= \frac{deg(a)}{m-1}= \frac{\sum_{u=1}^{m} x(a,u)}{m-1} \ a,u =1,2,3,...,m.
\end{equation}
In Eq. (1), the neighbors that are directly connected to node $a$ are represented by $deg(a)$. The $D(Ca)$ explains the proportion of a link from $a$ to $m$ among the real connected nodes and divided by $m-1$, which shows the majority of connections. To determine the limit of the most active node, high-degree centrality holds a bigger quantity of associations among the other nodes in the network. To exchange information among all nodes, a central node holding the locality mediates the defined network environment.

\subsubsection{\textbf{Betweenness Centrality}}
\par Betweenness centrality measures the amount of control over the information of a node flowing among other nodes. It shows the ties and locations of nodes. The shortest path, $P_{lm}(s_b)$, is calculated using the $s_l$, and $s_m$ that involve node $s_b$. However, $P_{lm}$ is the overall shortest path between $s_l$ and $s_m$. The following equation represents betweenness centrality:

\begin{equation}
B_c(s_b)=\sum_{l=1}^{n} \sum_{m=1}^{l-1}\frac{P_{lm}(s_b)}{P_{lm}}
\end{equation}
\par The nodes having a high betweenness centrality can communicate with non-adjacent users and may affect the flow of data between social communities. Hence, betweenness centrality is an important metric, considered as links to be determined among communities.

\subsubsection{\textbf{Closeness Centrality}}
\par Closeness centrality is supportive in minimizing network congestion, and delivers the reciprocal of the mean geodesic distance among nodes. Since the closeness centrality demonstrates that a source node is nearer to receiver neighbors, it ultimately shows a reduction of stay time in the queue for the data packet. The shortest path is measured by $S_b$ including neighbor nodes within its approachable premises. Information is also provided by this social property on the time needed to spread information to all nodes via $S_b$. The closeness centrality of an $S_b$ node is provided by the following formula, and $n$ provides the approachable nodes in the ASNET. To normalize the value of closeness centrality (Cc) between 0 and 1, $n-1$ is divided by the total distance to all connected neighbors, $d_s(S_b,S_c)$, which is represented in Eq. (3) as given below:

\begin{equation}
C_c(S_b,S_c)= \frac{n-1}{\sum_{c=1}^{n}d_s(S_b,S_c)}
\end{equation}

\subsection{Similarity}
\par According to the concepts of sociology, a social network consists of a higher degree of transitivity. It means there is a high probability of two individuals being aware of each other if they have greater-than-zero awareness in common. The common-awareness technique is known as clustering \cite{32}. Similarities can be calculated with a common-link ratio, which is a concept that shows the nodes in a group having common interests. Nodes having the highest similarity have more chances to meet each other and share data. Highest similarity nodes can be more suitable for information distribution among a cluster of nodes. Thus, to calculate the similar-interest profile with an efficient method and with lower overhead, Li et al. \cite{33} presented a semantics matching technique. In it, $l_x$ and $l_y$ are denoted as the profiles of the nodes in Eq. (4), which calculates the similarity between two nodes as follows: 

\begin{equation}
Sim(l_x,l_y) = \frac{\sum_{u=1}^{z} \underset{k\in[1,f]}{\max} Sim(dx_h,dy_k)}{z}. 
\end{equation}

\par As shown in Eq. (4), the number of concepts is represented by $z$ in profile $l_x$, while $f$ denotes the quantity of concepts in $l_y$. The $Sim(l_x,l_y)$ compares a user defined similarity threshold $(0 < threshold \leq 1)$. If the $Sim(l_x,l_y) > threshold$, then $l_x$ is semantically related to $l_y$, and if $Sim(l_x,l_y) \leq threshold$, it demonstrates that $l_x$ has a low likelihood of being semantically related to $l_y$. $Sim(l_x,l_y) = threshold$ demonstrates that $l_x$ is equally related to $l_y$. There must be a greater than 0 threshold value of social similarities to achieve the similarity level.

\subsection{Social Graph}
\par In a social graph, the related structure of social networks and trends is represented in a natural fashion. The edges in this graph characterize the individual social associations, and the vertices represent the human individual. The major challenge in a social network is the differentiation of edges between two vertices \cite{34}. In the more recent literature, several social graphs (e.g., the contact graph, the regularity graph, the neighbor graph, and the interest graph \cite{32,35}) are presented as having different edge meanings. These  graphs are all helpful in calculating the social levels of nodes, and they provide the easiness in communicating among nodes. To calculate the social graphs in the ASNETs,  consideration of node time, connection, frequency, and similarities are major concerns.

\subsection{Community}
\par Community represents the structure sub-unit that has a high density of internal links. Inside a community, users have more social connections with other individuals, in comparison to those outside the community \cite{34}. The social graph decides the social connections, which may be common interests, friends, family, and locations. Inside the community, individuals may meet frequently, due to their common behaviors. Thus, the mobility patterns of people are affected by the structure of the community, and they are also helpful in selecting a relevant forwarding path.

\section{Specifications of Congestion Control Protocol}
\par To achieve a high quality of service (QoS) in a congestion control protocol, the system should follow standards. Standardization of the system is helpful to achieve efficiency, compatibility, and interoperability in ASNET protocols. Some specifications are defined in the following subsections. 

\subsection{Efficient Utilization of Resources}
\par The performance of a network depends on the throughput of the system. Therefore, the design of a protocol should be to utilize the resources of the network efficiently within a limited time duration. To avoid wasting resources, the affected flow assigns its own bandwidth to the non-affected flow. Furthermore, for better utilization of resources and enhancing the performance of a congestion control protocol in an ASNET, the maximum bandwidth can be assigned to the node that has the maximum availability of data.

\subsection{Less Delay in a Queue}
\par Congestion control policies can improve the performance of a node if the queued packets are smaller than the buffer size. The small size of a queued packet can also provide maximum chances for other flows in transferring data packets. The size of the buffer needs to be reduced in order to reduce the queuing delay, but this technique suffers from low link utilization and can introduce a high packet loss rate. Therefore, for proper queue management in ASNETs, earlier transfer of data packets is required, and therefore, it is helpful to design a scheme that creates less of a delay in the queue.    

\subsection{Efficient Detection}
\par The methodology of congestion control should be able to detect different kinds of loss. This is because, in wireless environments, other than congestion loss, random losses can also occur that can affect the performance of a system from less utilization of resources. The degradation in resource utilization occurs when traditional congestion control modules are applied to ASNETs, e.g., traditional TCP always reduces the rate of a congestion window when it encounters any type of loss \cite{36}. In order to increase the throughput of the system, reducing the waste of resources and ensuring accurate detection of losses is compulsory.      

\subsection{Reduction of Buffer Overflow} 
\par Overflow in a node buffer can be reduced through the congestion control module. In order to reduce buffer overflow, each node should record its neighbor nodes' capacities. The capacity of a neighbor node provides an advantage in terms of reducing the rate of data on the sender side. Furthermore, the effectiveness of a node buffer can be improved by dropping some of the data from the node buffer and scheduling prioritized data packets first. The reason is that the aim of ASNETs is to provide maximum availability of prioritized data packets. Therefore, dropping-and-scheduling schemes can also enhance the performance of a node buffer by dropping useless data from the node and transferring prioritized packets earlier.

\subsection{Less Complexity/Overhead}
\par The design of a protocol should be simple and easy to implement. To avoid complexity in a system, the design should ideally not violate the layering concept. The information from other layers or cross layers can complicate the system and slow its performance. In an ASNET, where usage of multiple applications affects the performance of the network, a cross-layering concept degrades the system performance. Additionally, the system should reduce unnecessary acknowledgments to avoid overhead in ASNETs.

\section{Congestion Control Solutions}
\par In the modern era, after utilizing fewer resources, users want to communicate or perform social activities with other nodes ubiquitously, even if access to the Internet is not available. Therefore, in order to cover up the above-mentioned problems, ad-hoc networks require social properties that provide social communications everywhere, and that help in designing the ASNET. Social metadata and application data are mainly used in ASNET communications. However, the usage of social metadata and application data creates overhead on the scarce wireless resource. Moreover, due to the utilization of several social applications simultaneously at a single node and the large number of users in a single social community, the proper sharing of scare wireless resource among network nodes and avoidance of congestion-related losses is a vital task.
 
\par To resolve such issues, existing works provide congestion-related solutions without consideration of the socially-aware properties of nodes. Additionally, due to the shortfall in congestion control mechanisms for ASNETs, this area requires further growth. Therefore, in this section, we revisit the already developed algorithms/protocols for dealing with congestion problems. To understand the concept of congestion control in ASNETs, first, we study the existing work on congestion in networks, and secondly, we try to address how we can contribute to the design of a congestion control mechanism for ASNETs. 

\par To cover the above concepts, this section is divided into three sub-sections to discuss the existing congestion solutions. The first two sub-sections provide existing congestion solutions in ad-hoc and opportunistic networks. Additionally, to design protocols efficiently, in the last sub-section, we review some of the already developed bio-inspired models. We also summarized in Table 2 all categories with existing congestion control approaches. Therefore, we visit traditional ad-hoc and opportunistic networks to highlight the congestion issues for ASNETs. Additionally, limited work is available that uses social properties for communications in opportunistic networks, as indicated in Table 2. However, the workings of an opportunistic network are a little different than ASNETs. Opportunistic networks work only in a connectionless environment, however, ASNETs can work in both environments.  

\begin{figure*}
 \centering  
 \includegraphics[width=1.13\textwidth]{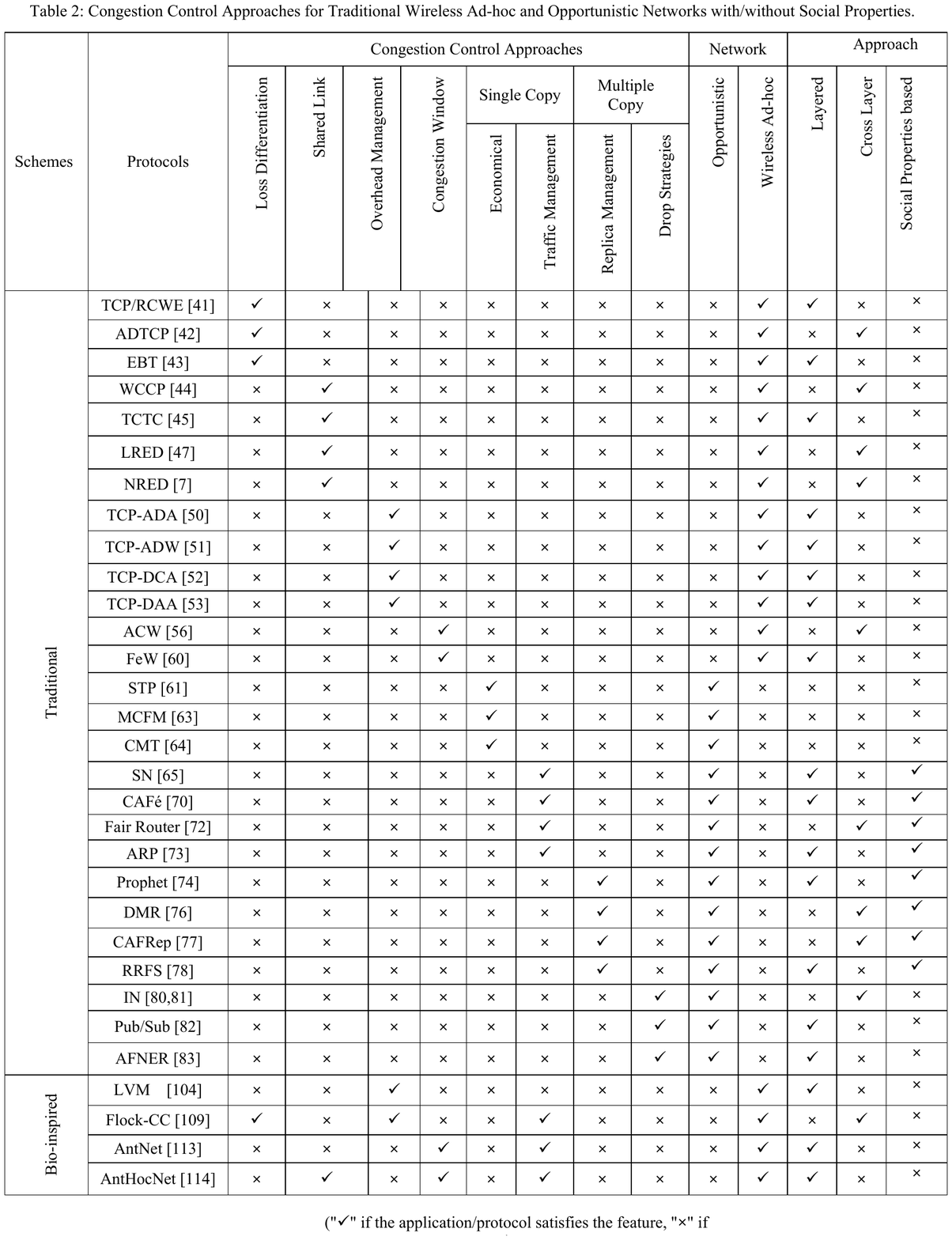}   
 \end{figure*}

\subsection{Ad-hoc Network Schemes}
\par In order to resolve congestion in a network, some upper-layer techniques were developed to alleviate and manage congestion without modifying transport protocol standards. Existing methods are mostly applied on top of an already developed transport protocol with the implementation typically residing at the application layer or in the operating system (OS) kernel. The implementations of these protocols are easy to deploy at the transport layer. Due to scarce bandwidth and wireless link losses, the above techniques cannot provide maximum data rates in ad-hoc environments. The reason for scarce bandwidth in ad-hoc networks is the shared resources, where every node competes to utilize the maximum bandwidth from the network. Therefore, to avoid congestion-related losses in ad-hoc networks, it is necessary to provide proper and efficient management of the data rate. 

\par In ad-hoc networks, the performance of the whole network is affected, not just a single node. To avoid congestion-related loss, TCP is considered the most prominent solution in wired environments \cite{36}. However, the performance of TCP is not good in ad-hoc networks. The low efficiency of TCP in an ad-hoc network is based on misunderstanding the conditions. This follows from the observation that congestion alone does not cause packet loss, which may also result from node mobility or wireless link errors. The non-efficient detection of loss from a lower round-trip time (RTT), makes TCP vulnerable in ad-hoc networks. Another drawback is that TCP-based senders wait for every acknowledgment. Waiting for an acknowledgment to every packet slows down the performance of an ad-hoc network. Furthermore, unnecessary acknowledgments put an ad-hoc network into a congested state at an earlier stage. Although existing TCP methods use a standard delay in acknowledgment options to reduce overhead in the network \cite{37}, it is not suitable for wireless ad-hoc networks due to the reliability issue. Moreover, congestion can be avoided at intermediate nodes after using the drop tail and random early discard (RED) approaches proposed by Kumar et al. \cite{38, 39}. The disadvantage of the drop tail method is that it drops packets when the size of the queue is greater than a maximum threshold. 

\subsubsection{\textbf{Congestion Handling using Loss Differentiation}}
\par In an ad-hoc environment, misinterpretation of losses can influence the performance of the network and lessen the utilization of resources. The reason for performance degradation in an ad-hoc network is less information at the upper layers. On the other hand, lower layers can perform efficiently in ad-hoc networks after providing reliability per hop. Then, the request to send (RTS)/clear to send (CTS) handshake and transmission attempts are used to decide whether to drop or sustain the packets. However, this method works only for the single-hop scenario. 

\par In an ad-hoc network, losses can occur not only due to a queue overflow but also due to movement of nodes or failure of a link. Providing end-to-end reliability with the congestion handling schemes in this sub-section, we provide a detailed analysis of existing schemes that will make it easier to understand the concept of the ad-hoc network. Moreover, we discuss some techniques to explain network information and to differentiate between wireless loss and wired loss. More detailed discussions of existing approaches are available \cite{40}. Additionally, in ASNETs, congestion handling using loss differentiation is a little bit different from traditional ad-hoc networks. This is because, other than the above-mentioned conditions, the reasons for losses in ASNETs may be due to the selfish behavior of nodes. In this selfish behavior, nodes usually do not want to communicate with nodes that do not provide any kind of incentive. Therefore, this factor also needs to be considered when designing a congestion handling protocol for differentiation of losses.   

\begin{figure}[t!]
 \centering  
 \includegraphics[width=.48\textwidth]{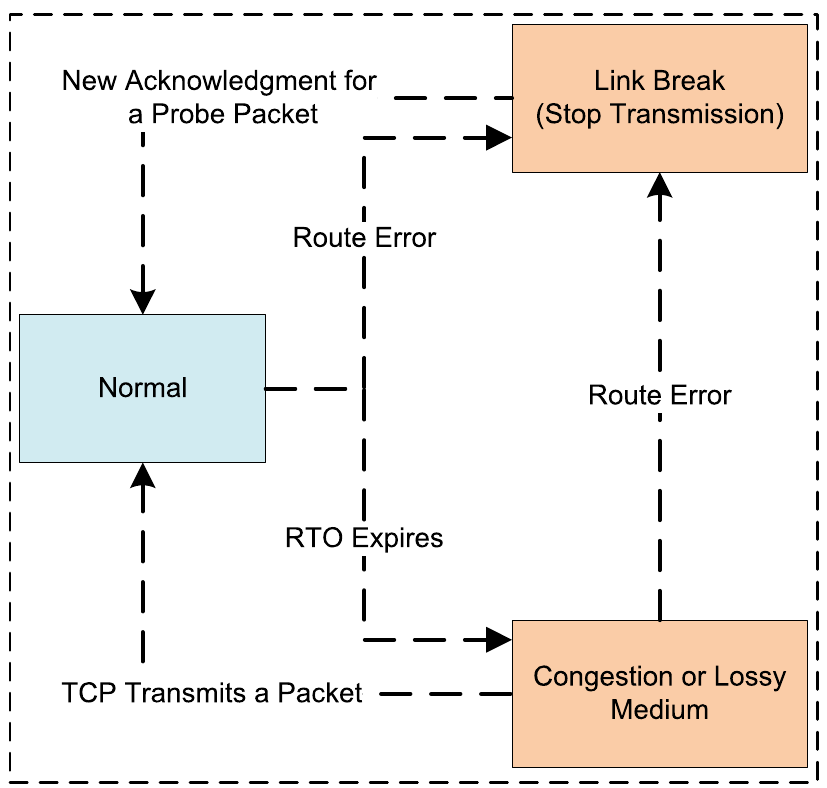}   
 \caption{Flow diagram of restricted congestion window}
 \vspace{-1.5em}
 \label{7}
 \end{figure}
 
\par To utilize resources efficiently, Gunes and Vlahovic \cite{41} introduced a TCP restricted congestion window enlargement (TCP/RCWE) solution that employs explicit information. It utilizes explicit link failure notification (ELFN) to detect the reasons for random losses and responds according to the situation in the network. The TCP/RCWE combines ELFN and retransmission timeout (RTO), which is helpful in detection and for taking proper action. The TCP/RCWE increases the congestion window based on traditional TCP, whereas the value of RTO is static. Additionally, if the value of RTO is increased, then it would not increase the congestion window. Figure~\ref{7} illustrates a flow diagram for the TCP/RCWE scheme. The performance of the TCP/RCWE shows better results in terms of higher goodput, such as 8\% and a loss rate is lesser, such as 14\% in ad-hoc wireless networks. However, the length of the congestion window is not too much, so the utilization of full resources can be affected by it. 

\par Fu et al. \cite{42} presented an ad-hoc transmission control protocol (ADTCP) technique for improvement in detection of the bit error loss rate. The ADTCP observed that in an ad-hoc network when an end-to-end approach is used, nosiness affects the performance of the network. Hence, to detect network congestion, ADTCP uses the difference in inter-packet delay at the receiver. Nevertheless, to estimate the number of packets in some specific time interval, it uses the metric throughput over the short term. In order to work in ad-hoc environment, the throughput of ADTCP increases from 312 Kbps to 567 Kbps (more than 80\%). Furthermore, ADTCP detects channel errors and route changes through the packet loss ratio and the arrival of out-of-order packets in ad-hoc networks. The sender uses the feedback from  the most recent network situation from the receiver and works properly with standard TCP. To distinguish between congestion and link loss, de Oliveira et al. \cite{43} presented an edge-based technique. They distinguished between losses using RTT, and did not utilize congestion control when a loss occurred due to a wireless link. In addition, their technique also uses RTT to detect route failure.  

\subsubsection{\textbf{Shared Link Congestion Control}}
The contention issue occurs, when multiple nodes use a shared link. Therefore, Zhai et al. \cite{44} produced a scheme called the wireless congestion control protocol (WCCP). The authors argued that TCP unfairness occurs due to contention, so WCCP calculates the network congestion level after considering the channel usage ratio. WCCP provides a cross-layer solution that is based on the transport layer and the layer between the network and media access control (MAC) layers. Conversely, it assigns resources to inter-nodes and intra-nodes after estimating the shared channel's available bandwidth. Besides, Hamadani and Rakocevic \cite{45} presented TCP contention control (TCTC) to solve the problems of uncertainty in the TCP stream, which is affected by unnecessary transmission of data. TCTC considers the dropping of data packets that occurs through excessive transmission of data in a single flow. For a fixed probe interval, TCTC achieved  average 34\% improvement in throughput, and the 40\% decrease in delay packets is caused by the level of contention. These methods are helpful in estimating congestion on the receiver side and informing senders. Moreover, to calculate the optimal amount of traffic, TCTC also provides a new TCP contention window.

\par In addition, Hongqiang et al. \cite{46} designed another method that is based on two systems. First, it designs the current receiver with a high priority that accesses the medium in order to prevent within-path contention for every user. Secondly, based on the backward pressure scheduling method, the authors assigned limitations on the transferring user when delivering extra packets to a downstream user, even though that downstream user can accept further packets for a specific flow. For TCP that attains the optimal performance, Zhenghua et al. \cite{47} proposed an efficient congestion window. Despite discovering the efficient congestion window rate, TCP raises its congestion window aggressively, which can drop packets at the data link layer. Still, the authors discussed the reason for packet drops in wireless networks as being infrequent buffer overflows since a wireless network provides a buffer size greater than 10 packets. The major cause of packet drops in wireless networks is medium contention. As a solution, loss ratio-based random early detection (LRED) and an adaptive pacing design address the unfairness and contention problems. The LRED scheme discusses medium access that retains the average number of attempts. The scheme specifies that when the value is greater than a specific average threshold, then the calculation for the dropping probability of packets is based on the RED algorithm \cite{48}. In the RED algorithm, dropped packets notify the sender that is creating congestion on the network. But, in the LRED scheme, the dropped packets indicate that it is nearing an increment in the contention level in the network. Furthermore, LRED uses the adaptive pacing method that is used for adaptation of data packet flow in an additional fairness approach. The drawback with LRED is, that it slows down the flow of data packets, and utilizes the resources of the network poorly.

\par In a scheme that addresses the issue of unfairness in the wireless medium, Xu et al. \cite{xu2005tcp} designed a neighborhood random early detection (NRED). The performance of NRED and the fairness among the nodes is enhanced in a wireless environment after depending on the RED algorithm. The NRED algorithm designed a distributed neighborhood method based on a collection of the neighborhood queues' length information. A node estimates the distributed neighborhood queue size, and it calculates the probability of guaranteeing fairness with dropped and marked packets. This scheme is data link layer dependent, and can be divided into three parts: 1) distributed neighborhood packet drop (DNPD); 2) neighborhood congestion detection (NCD); and 3) neighborhood congestion notification (NCN). NCD calculates the distributed queue using the average queue size. NCN investigates the utilization of the channel for multiple time slots and determines what time, and with which method, a node should inform its neighbor nodes about congestion. Besides, the DNPD calculation is based on the drop probability of the local node.

\par Additionally, the shared link congestion control operation in ASNETs is comparatively different from traditional ad-hoc networks. Transference of huge amounts of data by some popular nodes leads to congestion. That is why, if there is a delay in transferring data packets from popular nodes to an intermediate node, then congestion occurs. However, construction of a packet scheduling algorithm that operates according to the concept of ASNETs is a solution for reducing congestion on intermediate nodes. The consequence of this packet scheduling is the sharing of bandwidth resources among nodes. Moreover, it also provides a solution to various problems caused by multiple connections after sharing one link.
\begin{figure}[t!]
 \centering  
 \includegraphics[width=.49\textwidth]{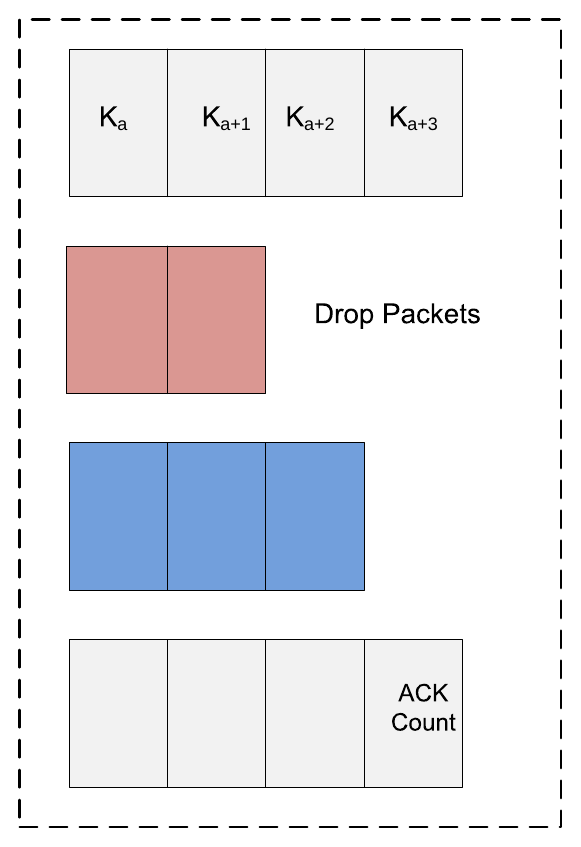}   
 \caption{TCP dynamic adaptive acknowledgment delay window adjustment}
 \vspace{-1.5em}
 \label{8}
 \end{figure}
\subsubsection{\textbf{Acknowledgment Overhead Management}}
For a reduction in traffic overhead, this sub-section provides a discussion of existing methodologies for delaying acknowledgment. In this paper, we also discuss the drawbacks in the existing methodologies and efficient solutions for ad-hoc networks. The reason for performance degradation in an ad-hoc network is due to an increment in overhead from acknowledgment packets. This is because data and acknowledgment packets use the same path for communication, and acknowledgment packets compete with data packets, which can cause collisions in ad-hoc networks and deplete the resources earlier \cite{49}. Consequently, some segments in ad-hoc networks should be reduced to overcome reduced performance of the network. To overcome the above issues, a lot of work is available on reducing the data segment, and more work is available on reducing spatial contention, which is done by minimizing the acknowledgment segments. A reduction in acknowledgments can be provided through a delaying mechanism that sends a cumulative acknowledgment, rather than many single acknowledgments. Nevertheless, the usage of a cumulative acknowledgment can degrade the performance of TCP because when an acknowledgment is not received in time, TCP assumes packet loss has occurred. In the following sub-sections, we discuss existing approaches that were developed for wireless networks as a result, but they cannot work properly in ASNETs. This is because in ASNETs, some sender nodes are popular and demand rapid acknowledgment for reliable and early communications.

\par For providing a solution in mobile ad-hoc networks (MANETs), TCP with adaptive delayed acknowledgment (TCP-ADA) \cite{50} provides a delay in-acknowledgment methodology on the receiver side. TCP-ADA sends a single acknowledgment after receiving all data packets from the full congestion window. After the reduction in the acknowledgment number, spatial contention can also improve throughput effectively. In comparison to a simple delayed acknowledgment scheme, TCP-ADA provides 30\% improvement in throughput. On the other hand, in TCP-ADA, some drawbacks reduce the performance of the network. The first drawback arises when any acknowledgment packet loss occurs in an ad-hoc network; then, it is difficult to regain performance. This is because the lost acknowledgment packet cannot be recovered easily after an adjustment for a new congestion window. A second issue arises in terms of RTO when the sender will resend the data packet if it does not receive or recover the acknowledgment packet within a certain time. However, this paper has not addressed the performance of a network in a dynamic environment. Al-Jubari and Othman \cite{51} provided a solution for overhead control in static networks, called TCP with an adaptive delayed-ACK window (TCP-ADW). The work of TCP-ADW is also based on the receiver side and shows that the static delay window size is not good at achieving high throughput. Hence, it uses a dynamic window size based on the arrival time of data packets on the receiver. In TCP-ADW, the receiver side always increases the delay window unless RTO or the session starts up. Compared to traditional TCP, the performance evaluation of TCP-ADW increases 233\% throughput in multi-hop wireless network. Nevertheless, TCP-ADW cannot perform efficiently in ad-hoc networks after considering the inter-arrival time of packets only. This is because a delay of packets in an ad-hoc network cannot only be due to congestion but also due to wireless link loss. 
\par A TCP with delayed cumulative acknowledgment (TCP-DCA) \cite{52} specified that when a delay acknowledgment technique is applied, TCP cannot gain higher throughput. Therefore, to enhance TCP throughput, it uses the concept of an optimal size for the delay window at the destination node. The authors proposed that path length is important when selecting the optimal size of the delay window. Hence, TCP-DCA uses the number of hops to select the delay acknowledgment window size. TCP-DCA performance is better in multi-hop ad-hoc networks and shows that if the amount of data is large, the larger delayed window is beneficial in a shorter path, but it may work inversely if the path is longer. Although TCP-DCA also has some drawbacks when the end-to-end path is longer, it is difficult for TCP to find the reason for packet loss in a longer path. TCP-DCA also faces problems due to channel performance with a high bit error rate because this technique uses a large fixed size (four) for the delay window.

\par Finally, Oliveira and Braun \cite{53} presented a delayed acknowledgment methodology called TCP with dynamic adaptive acknowledgment (TCP-DAA). To reduce acknowledgment overhead, TCP-DAA is based on both sender and receiver. In TCP-DAA, when the status of a wireless channel is better, it sends one cumulative acknowledgment after receiving four data packets. TCP-DAA provides an advantage for reducing collision loss after setting a minimum congestion window limit on the sender side, and it performs efficiently in multi-hop conditions (approximately 10 hops). TCP-DAA works in a dynamic environment and delays acknowledgment until it receives four data packets. The setting of delay acknowledgment depends on the condition of the channel. If the channel is in good condition, the receiver replies to the acknowledgment earlier; otherwise, TCP-DAA sets the delay window dynamically. It also reduces the size of the acknowledgment window when TCP-DAA faces any loss. However, in comparison to existing TCP, the rate of data in TCP-DAA is 280 kbps. Figure~\ref{8} illustrates the delay window adjustment method of acknowledgment. Nevertheless, this technique also has some drawbacks because of the fixed size of the delay window (i.e., because of this, it cannot work better at more than 10 hops). Although a lot of schemes are designed for ad-hoc networks, none has explained the social-awareness and proper consideration of network conditions for variable adjustments in the delay acknowledgment window. To search for similar interests, two types of data are generally used in ASNET communications: 1) social metadata, and 2) application data. However, the bandwidth is a scarce resource and the usage of social metadata and application data may cause network congestion. Therefore, other than the traditional overhead control schemes, there is also a need to control social metadata that creates overhead in ASNETs.

\subsubsection{\textbf{Congestion Window Adjustment Methods}}
\par The adjustment of the congestion window depends on the situation in the network and provides a great advantage in the avoidance of congestion-related loss. For that reason, in this sub-section, we discuss some existing approaches that are helpful in adjusting the rate of the congestion window. Rangwala et al. \cite{54} proposed a rate control adjustment approach that uses additive-increase and multiplicative-decrease methods. The adjustment in rate control is variable and depends on the estimated available bandwidth of the network. Therefore, the estimation of the bandwidth is very important for this method type and needs to be highly accurate with reasonable bandwidth allocation. This work is helpful in congestion handling and avoids wasting resources. Similar to above, another work presented by Triantafyllidou et al. \cite{55} adjusts the congestion window for the bandwidth delay product (BDP). This scheme also adjusts the maximum congestion length based on a time expression. The calculation of BDP in a wired network depends on the multiplication of RTT and the bandwidth. Nevertheless, the calculation of BDP in a wireless network is based on the density of nodes, the power of transmissions and the buffer capacities. The calculation method in a real BDP network in a complex contention method is difficult. 

\par In recent work, Hamadani and Rakocevic \cite{56} proposed the adjustment of the congestion window that controls contention through TCP. To estimate the status of contention through the path and the level of throughput, the authors presented two distinct parameters. Furthermore, these two parameters are used to adjust the congestion window after combining four possible results. The variable congestion window adjustments and the estimation method create complications and do not provide accurate values to estimate the network condition. Zhang et al. \cite{57} also used a TCP contention control method already defined by Hamadani and Rakocevic \cite{56}. Like Hamadani and Rakocevic, Zhang et al. used an estimation method that is dependent on throughput and contention. These methods do not provide efficient results due to the lack of information about the whole network and can only inform about single-node congestion status. However, as compared to TCP and its contention control methods, TCPCC gets 750 packets in terms of the throughput and 0.2 seconds for the end-to-end delay.

\begin{figure*}[t!]
 \centering  
 \includegraphics[width=1\textwidth]{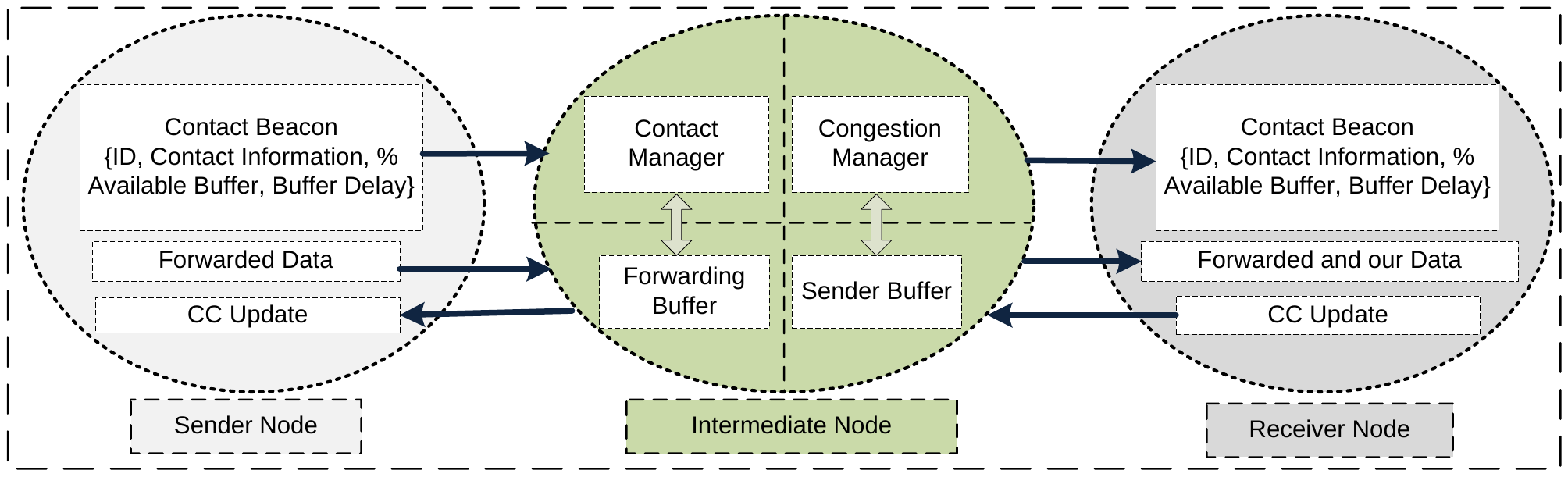}   
 \caption{Architectural design of the congestion-aware forwarding algorithm}
 \vspace{-1.5em}
 \label{9}
 \end{figure*}
 
\par In a wired network, the adjustment of the congestion window uses the traditional TCP method to resolve the congestion issue. However, the traditional TCP method cannot provide efficient results in wireless ad-hoc networks. Therefore, many researchers have done a lot of work to improve traditional TCP to work in a wireless ad-hoc environment, on things like increases in the congestion window at a certain fractional rate. Nahm et al. \cite{58} proposed some methods that make a decision after each RTT. To avoid the occurrence of congestion in the network, slow congestion avoidance (SCA) scheme was proposed by Papanastasious and Ould \cite{59}. The traditional methods increase a congestion window linearly that is, the maximum segment size (MSS) after receiving acknowledgment for all data packets. 

\par To reduce the overshooting issue in the congestion window, Nahm et al. \cite{60} provided another solution that uses fractional window (FeW) increment concept. Calculation of the congestion window is based on the sent data packets and received data packets within one RTT delay. The aforementioned method provides a slow increment in the congestion window. FeW performs efficiently after the slow adjustment of the congestion window. It improves the performance of TCP and network stability in multi-hop networks. The performance of the system is improved by 90\% compared to the traditional scheme. On the other hand, it is improved around 70\% compared to the existing variants of TCP (ADTCP, LRED). Additionally, working congestion window adjustments in ASNETs is different, compared to traditional ad-hoc networks. In ASNETs, people often communicate and share maximum resources with those nodes that have a maximum similarity index level. The assignment or adjustment of resources based on social interest is a key factor for ASNETs. Therefore, this factor also needs to be considered with the existing methods when managing resources and reducing congestion loss.

\subsection{Opportunistic Network Schemes}
Communication between nodes in opportunistic networks does not depend on end-to-end connectivity. The reason for non-availability of nodes for end-to-end communication is due to longer delay among the nodes. Based on the reduced availability of the end nodes, the protocols designed for traditional ad-hoc networks cannot work properly in opportunistic networks. In addition, longer delays and the non-availability of end nodes also create hurdles in  identifying and rectify the problems. Consequently, in opportunistic networks, nodes face reliability and congestion issues. To adjust for the problem of congestion in this paper, we did a literature review related to congestion control schemes that identified how congestion occurs and which solutions are available in opportunistic networks. Furthermore, in opportunistic networks, some work is also available that uses socially-aware techniques to control the congestion issue. Nevertheless, these techniques solve congestion issues through routing methods, and cannot work in traditional ad-hoc networks. In order to clarify the workings of ASNETs by using opportunistic networks, further social properties should be included. The management of congestion traffic through a single-copy is a desirable factor in ASNETs. This is because a single-copy management system provides efficient resource utilization. Furthermore, after using social properties, management of replicas, and dropping strategies, using multiple copies also plays a vital role in ASNETs. To solve congestion issues in opportunistic networks, the following sub-sections provide a detailed discussion of the existing methods.  

\subsubsection{\textbf{Economical Single-Copy Congestion Management}}
\par Message storage management through a single-copy strategy is helpful in solving  congestion issues in nodes because the buffers of mobile nodes have a limited storage capacity. Thus, to resolve such an issue in an opportunistic network, a relay node drops messages from its own buffer after they are transferred to the next node. However, if a forwarded message is dropped for some reason, then this strategy cannot recover that message due to unconnected nature of the source node. Consequently, this strategy degrades the delivery ratio performance in the network. To discuss such problems in opportunistic networks, we did a literature review in storage management policies that use economic models. The maximum existing models use local information about nodes because of the difficulty in obtaining global information in an opportunistic network. 

\par To transfer messages within a short period of time and to lessen consumption of the relay node buffer, Fall et al. \cite{61} provided an efficient economical decision model for the acceptance of messages. The selection of messages based on a shorter time period gives an advantage in earlier buffer release. Still, a smaller size for the message is also helpful for the avoidance of congestion in a node buffer. To improve the message selection method, further properties like message priority, security, routing strategy, and the message lifetime can also offer advantages in a network. Furthermore, the delivery performance and the utility of resources can be improved after applying the game theory mathematical model \cite{62}.

\par An acceptance decision for message custody based on a financial model was presented by Burleigh et al. \cite{63}. In this model, delivery charges for the messages are paid by the sender, and the relay node charges a commission for transferring the message to the next node. The authors defined delivery charges with a message size function and a request for QoS. Because of the incentive for the relay node, it quickly transfers larger messages. The drawback to this scheme is that if the relay node receives a message larger than the remaining time, that message can create congestion in the node buffer. Consequently, the remaining lifetime of a message and the growth rate of the queue are the main factors for the acceptance and the rejection of a message.   

\par To design a dynamic opportunity cost in a delay-tolerant network, Zhang et al. \cite{64} presented a method of message storage with congestion management techniques. The proposed design uses dynamic programming and the revenue management concept. The forwarding of a message is based on the assumption of a node's minimally cooperative behavior, and the overall revenue is optimized through a supervised transfer of the message. Acceptance of the message in this design depends on two demands: the opportunity cost and the benefit function. The opportunity cost is based on the storage capacity value that has been consumed. Furthermore, in opportunistic networks, the function of message type or size is denoted by the benefit function. The proposed scheme utilizes 80\% resources with the avoidance of congestion and load balancing. All the aforementioned methods provide an economic model after using the single-copy concept in opportunistic networks.

\subsubsection{\textbf{Traffic Management Using a Single-Copy}}
\par In social opportunistic networks, where human-to-human or node-to-node communication takes place, a significant problem can arise from biased resource allocation. Connections among devices are only occasionally available in opportunistic networks \cite{ahmed2014event}. Besides, few devices being handheld devices are dependent on the movement of humans in the network. This sub-section looks at the literature covering degree, betweenness, closeness, and ego betweenness centrality related to the intricate social activities of the nodes. Centrality-based social properties are cooperative in the management of data storage and traffic distribution. 

\par According to Hossmann et al. \cite{65}, the structure of social networks is non-random. Some nodes work as a communication hub in this non-random behavior. Such a structure exhibits a few nodes that work as a communication hub and keep most of the network data. Therefore, the hub nodes are selected based on higher popularity or the highest centrality in the social network. For better transmission of the message, Bubble Rap \cite{66} and SimBet \cite{67} utilize a relay node to the hub node and are prominent examples of socially-aware routing protocols. According to experimental results, fewer deliveries of data packets can be caused by the unfair distribution of the load and can be the reason for congestion at central hub nodes. The experimental results showed that to increase delivery performance, SimBet transferred 9022 messages over the network. Additionally, the Bubble Rap achieved 80\% success in delivery ratio. Kathiravelu et al. \cite{68} as well Khabbaz et al. \cite{69} demonstrated that the hub node is superlative for forwarding data because of its central site in the whole network and the ability to receive all data from the other nodes. The above two solutions delivered answers using the central node for management of the storage space in order to clarify the congestion problem. The sender node must forward a message to get the status regarding the level of congestion prior to the submission of data. 

\par Radenkovic and Grundy \cite{70} presented congestion aware forwarding (CAFe) to lighten the burden on the hub node. CAFe works using two main components: congestion control and routing. The CAFe routing algorithm depends on social-based routing. Figure~\ref{9} illustrates the architectural design of the CAFe routing algorithm, which uses the ego betweenness centrality concept to transfer data among nodes. The highest betweenness centrality specifies a short path between the receiver and the sender. Thus, the transfer is performed without data waste. In an ideal scenario, for selection of the relay node, betweenness centrality is an optimal choice. This property needs all the information about the network. In opportunistic networks, such a situation is difficult to achieve. Thus, it is a disadvantage of the betweenness centrality property. Hence, this scheme uses an ego network \cite{71} to compute an approximate betweenness centrality. After assessing the measurements of the node buffer, the other module controls congestion. This component controls congestion by approximating the node buffer's statistics. The decision provided by the buffer level is either rejection or acceptance of the messages. The congestion information is difficult to calculate for the whole network in opportunistic networks, so CAFe computes the congestion level based on ego networks. 

\par In another relevant work focused on the distribution of traffic in social opportunistic networks, Soelistijanto and Howarth \cite{29} proved that a routing metric cannot perform appropriately with ego betweenness centrality. Their results showed that for faster distribution of traffic, the ego betweenness centrality gives poorer performance than degree centrality. In social networks, the solution to unfair load distribution was provided by Fair Route \cite{72}, which is similar to CAFe. The authors proposed a forwarding algorithm that depends upon routing and congestion. The congestion is controlled by the queue length of the buffer in the module for congestion control. A forwarding request or its acceptance is dependent upon the queue size of the node. The size of the queue determines the assignment level of the nodes (higher or lower). Tie strength can also represent the level of interaction which predicts future communications (i.e., strength is added with increments and decrements in node contact, exponentially over time). Nevertheless, considering only tie strength is a disadvantage. Consequently, statistics about the node buffer should be considered for a balanced distribution of traffic while forwarding a message. Therefore, FairRoute performed 33\% better than existing methods. Furthermore, SimBet achieved 80 Kb cost for the reader that is higher than FairRoute. However, as compared to SimBet, FairRoute achieved 20 Kb cost for the reader.

\par Kathiravelu et al. \cite{73} presented an adaptive routing protocol to calculate probability metrics by using the degree of connectivity among the nodes. This technique expresses the maximum connection among nodes, which shows the popularity of a node in the network. Then, the popularity of the node becomes a reason for the selection of a relay node. Therefore, the adaptive routing protocol achieved 90\% performance when the value of the connection is 50\%. Nevertheless, this scheme presents a drawback in terms of congestion that occurs because of the maximum number of links with relay nodes. To answer the above-mentioned congestion issue, the congestion aware adaptive was proposed by Kathiravelu et al. \cite{68}. This technique informs other nodes (using advertisement methods) about free buffer space availability. The degree of connectivity is calculated by every node with each neighbor node in the first step while, for computing a safety margin, the popularity level is used in the second step. Therefore, it achieved 80\% transfer of messages, when the size of the queue is 20 kb. 

\subsubsection{\textbf{Replica Management using Multiple Copies}}
In opportunistic networks, message replication helps to improve the average delivery ratio due to worse storage congestion at relay nodes. But, it is hard to attain the finest transfer efficiency by determining the precise quantity of message copies in opportunistic dynamic networks. Prophet \cite{74} and MaxProp \cite{75} are routing protocols that bound message replication with a definite forwarding policy. The dependence of the forwarding policy is based on the node delivery probability and contact history. However, the protocols replicate the message, irrespective of network congestion state, whenever a message reaches the criteria for replication in the buffer of a node. Hence, dynamic replication control should adaptively adjust the replication rate of the message to the level of the network. 

\par Thompson et al. \cite{76} employed dynamic message replication by retiring the replicant to limit the network congestion level's rate of replication. To learn the current level of network congestion, a model for congestion control was proposed that monitors the entire network and feeds the information back to all network nodes on time. A mathematical model spreads a sole message throughout the network and achieved 280\% delivery rates in the defined scenario. For network-wide congestion level measurement, proxy metrics are employed \cite{76}. In addition, the ratio of the buffer occupancy rate or message drop rate to either the receiving message rate from neighbor nodes or the end-to-end acknowledgment rate is a worthy sign to find the congestion level of the network. Simulation results show that at a low congestion level, node buffer occupancy is high, and acknowledgment spread is untrustworthy, whereas it gets deferred in opportunistic networks. Hence, it was concluded that the ratio of the receiving message rate to the message drop rate is the best local metric at a single node to represent the congestion level in the whole network. 

\par Current congestion information is exchanged by nodes during node contact, and they independently estimate the current local congestion level. Whenever, a node contacts other nodes, it updates its local estimated congestion and adjusts its replication limit. The congestion level decreases with gradual increases in the replication threshold. Subsequently, the congestion level increases when the replication threshold is reduced multiplicatively, which acts like TCP's additive increase multiple decrease (TCP-AIMD). In opportunistic networks, it is worth noting that message drops not only depend on storage congestion but also on node inter-contact time. As the lifetime of a buffer expires, dropped messages increase. Hence, it is difficult to accurately measure the message drop rate at the node congestion level. The selection of message lifetime should be carefully achieved because the message drop rate will be high for a shorter lifetime even though the buffer is not congested. 

\par A multiple-copy variant of CAFe \cite{70} called CAFRep \cite{77} was proposed to control the number of message copies at relay nodes. CAFRep has three important metrics: ego network statistics, node buffer statistics, and node social-network metrics. Total node utility (TotUtil) is calculated as the summation of the metrics of a node that applies CAFRep. A node compares its total utility, during node contact, with a contact peer that is used for selection of the next hop, and the number of messages is decided upon to determine which are to be copied at the node. In order to achieve efficient performance in congestion control methods, CARep achieved 70\%. Furthermore, as compared to CAFE, it provides 80\% availability of buffer space. Yin-Ki et al. \cite{78} proposed round robin forward scheduling (RRFS) for replication management that controls the sequence of the transmission of messages by following a routing policy. In opportunistic networks, provision of a high delivery ratio with prioritized forwarded messages is the main goal of routing algorithms based on message lifetime or message service class and message delivery probability. On the other hand, excessively replicated messages at the front of the queue are avoided by the RRFS algorithm through message prioritization. In the network, the total number of multiple replicas is estimated for every message. The message header information is stored whenever a message gets replicated during node contact, and the total number of multiple replicas is incremented. This procedure reduces the network congestion by efficiently controlling the message copies in the limited contact time. Additionally, the proposed methodology also reduced the delivery delay ratio up-to 10\% to 25\%.

\subsubsection{\textbf{Drop Strategies using Multiple Copies}}
In multidisciplinary networks, the buffer of the current node retains a copy of a message forwarded to another node. The node uses this copied message for future correspondence with other nodes. The matter of concern is that when the buffer reaches its storage capacity it starts randomly discarding the stored messages \cite{79}. Two strategies were adopted as abatement measures to the problem mentioned above. First is storing single-message statistics that merely include message attributes in the buffer of a node (e.g., message lifetime, arrival statistics, and forwarding). 
Second, keeping network-wide message statistics that include attributes of messages gathered from the entire network (e.g., the amount of message copies) is a relatively intricate drop strategy. The performance of the proposed system achieved 0.83\% delivery ratio and 519.75 sec delivery delay in the case of unlimited buffer and limited bandwidth. The design system also discussed the case where the bandwidth and the available buffer are limited. In the defined scenario, the system achieved 0.55\% delivery ratio with 1469.5 sec delivery delay. The strategy to drop packets in intermittent networks \cite{80, 81} provides simple policies for dropping messages. Two parameters are used to check the efficiency of a network that includes delivery delay and ratios. 

\par A pub/sub-based model was proposed by Bjurefors et al. \cite{82} to classify several drop probabilities of data-centric opportunistic networks. The Hub node in infinite buffer space get 83\% achievements in delivery ratio. With an average forwarding number based on epidemic routing \cite{83}, a drop policy for opportunistic networks is provided when the addition of new packets needs buffer storage while the buffer is fully occupied. This algorithm only works when the buffer is completely full and a new message arrives. Rather than observing the average forwarding count of the whole network, it randomly discards messages where the forwarding amount is higher. As compared to the traditional epidemic routing scheme, the proposed system attains 86.85\% delivery ratio when there were 100 packets. Additionally, the proposed scheme achieved 80 secs end-to-end delay when the network bandwidth is 2 Mbps. However, the global forwarding number is neglected in this study, which can be explored further. Additionally, an optimal buffer management drop strategy was proposed by Krifa et al. \cite{79}.

\subsection{Bio-inspired Schemes}
The communication structure of the Internet is increasing day by day due to a large number of nodes, which is also increasing the congestion levels in networks. The evolution of the Internet occurs at a fast rate, exploiting advanced technology to fulfill the latest demands. This evolution has proven to be a real challenge to the Internet's future and to our ability to solve different issues that we currently face, e.g., overall system security, congested environment, routing scalability, and networks pressurized by future devices and recent applications \cite{84}.  
\begin{figure}[t!]
 \centering  
 \includegraphics[width=.48\textwidth]{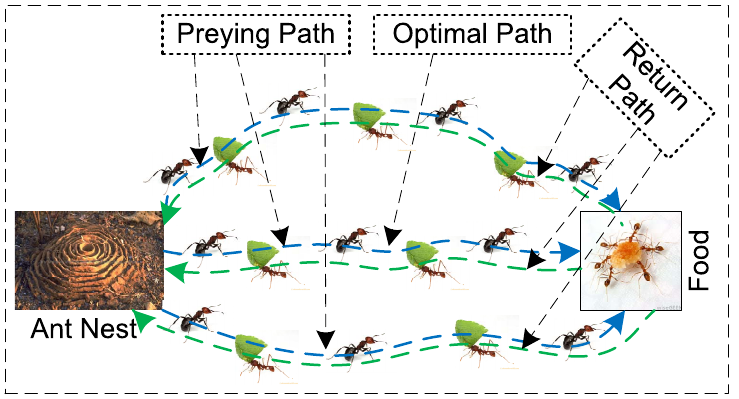}   
 \caption{The ACO-based optimal path selection}
 \vspace{-1.5em}
 \label{10}
 \end{figure}
\par In comparison with traditional central networking, distributed networking plays an important role in satisfying present and future Internet network design requirements. However, to design the requirements of a distributed system in an efficient way, some biological processes are too significant. Biological processes employ simple and identical operating rules for every individual to interact with others, producing efficient patterns toward the achievement of the whole system. For example, traditional routing schemes cannot provide efficient services due to the congestion created by messages and the data storage requirements of large routing tables \cite{85}. We can also take examples, such as better output by utilizing less energy and inducing modifications in different systems (i.e., the cloud, the data center, and the grid \cite{86}). It is not possible to align the parameters of each unit through a specific hierarchical control because the above-mentioned systems consist of scattered units. The organization of a system's operational achievements and quick power decisions against failures in other units can be rationalized by networking-based control that empowers the whole process.

\par A long time ago in communications technologies, there were new results from evolution, for example, congestion, heterogeneity, complexity, and scalability. The challenges from these new results can be dealt with gracefully and efficiently as works of art created by evolution. A clear observation shows that biological systems and rules depend upon some generic criteria with effective results, such as resource management, social differentiation, task assignment, and organizing systems without any external source (e.g., protection of the immune system without brain control \cite{87}). Additionally, the functions of organisms are performed in a balanced state that is related to homeostasis \cite{88}. Operating system resource management protocols for Internet of Things (IoT) are discussed in detail in \cite{152, 153}. In insect colonies, a task is decided on and executed as individual responses that create a global intelligent assignment \cite{89}. Bio-inspired systems and organizations have demanded features because of the long-time ongoing progress in evolution. Consequently, to elaborate on these features, some key points are given below.
\begin{enumerate}
\item The system should behave properly in a dynamic environment;
\item The recovery of the system is capable of solving the problems due to internal or external causes;
\item Adaptation of the system takes place upon the arrival of new conditions;
\item Dealing with unnatural resources is done with the help of a vast aptitude;
\item There is an ability to adapt with perfection, and gaining an effective balance;
\item Facing difficulties is accomplished due to available inherited abilities. 
\end{enumerate}

\par All of the above-mentioned features reflect the effectiveness of using bio-inspired techniques in conjunction with traditional technologies. Hence, many researchers are motivated to apply principals of biological methods in order to resolve the technical issues of networks \cite{90,91}. These biological methods introduce new procedures for communications systems and provide predictable technological systems. Therefore, to solve congestion issues in the Internet and ad-hoc environments we discuss the available bio-inspired techniques in following sub-sections. In addition, complete surveys of bio-inspired techniques have been published \cite{92,93}. The use of social networking in conjunction with bio-inspired approaches can lead to more efficient designs for ASNETs and should be explored in future work. 

\subsubsection{\textbf{Internet Environment-Based Congestion}}
A major cause of degradation in Internet performance is congestion. Consequently, to overcome the congestion issue through realistic solutions, bio-inspired techniques can provide advantages. In algorithm and structural design, the Internet and biological systems have many analogies. The Internet architecture is based on application, transport, network, data link, and physical layers. Just like the Internet, a biological system architecture also depends on other systems. There is always a resemblance to the infrastructure of biological systems. For example, many nutrients are food for bacteria, and these nutrients are responsible for energizing the bacterial cells. A bacterium digests these nutrients and rearranges the components of the nutrients into multifarious giant molecules. Because it is impossible for a bacterium to convert each nutrient into each macromolecule through metabolism, in the same way, it is not feasible to compose submissions for each physical layer technology. Bacterial metabolism renovates nutrients into some common prevalence, which is then used to compose intricate macromolecules for energizing the cell.

\par Csete and Doyle \cite{94} observed that this bow-tie structure is a unique component. Let us consider the Internet protocol stack, consisting of similarities at a single protocol level. A standard congestion control algorithm is the basis of TCP, and it also utilizes responses from the receiver's acknowledgment to adjust the sender's transmission rate. Conversely, that type of feedback arrives as positive and negative feedback. Initially, when senders acquire the desired acknowledgment (positive feedback), then there will be a gradual rise in communications. Secondly, when senders do not acquire the desired acknowledgments (negative feedback), communications rapidly decrease. In short, if the existing condition of the network is known, the algorithm will match the transmission rate of the sender to the receiver, which is an important part of feedback in control theory. The latest research has proven that bacteria use this feedback for moving with a specific velocity when attracted by chemicals and food in their surroundings \cite{95}.

\par In order to solve congestion issues in Internet communications, Analoui and Jamali \cite{96} presented a framework called bio-inspired congestion control (BICC). After inspiration from biological systems, the congestion control framework uses the predator-prey model that controls the level of a population. In their paper, the authors first provided a short review of the population control methodology, and then combined it with Internet congestion control. The congestion control scheme utilizes a bio-inspired mathematical model and provides efficient results in communications. The relationship between the predator-prey  model and the Internet traffic control model is as follows: if predators are absent, then the growth in prey will be exponential. The authors considered the species in the Internet ecosystem similar to the predator-prey model. Furthermore, if prey is absent then the number of predators will be reduced exponentially. The results of the framework are efficient in terms of congestion control and achieved 10.4146 Kbps throughput. In this framework, some drawbacks exist---e.g., BICC does not consider the different RTT and the short flows effect.

\par A further extension of the BICC algorithm was discussed \cite{97} leading to realistic bio-inspired congestion control (RBICC). That paper also used the naturalistic predator-prey model to solve the congestion issue in the network. In the previous BICC framework, the authors discussed some parameters that are helpful in achieving efficient performance after fair utilization of resources. The major drawback in BICC was its speed of convergence and stability, which cannot behave dynamically. Hence, the RBICC authors improved the dynamic performance and used the predator-prey naturalistic model. Other than the intrinsic methods of nature, RBICC provides a mathematical and theoretical framework that is helpful in the design and analysis of a system. Consequently, RBICC defined three more natural methods ( competition, predation, and parasites), which are helpful in controlling the population in a more efficient way. The level of congestion in the network can improve after a reduction in population size. In RBICC, the amount of population control is based on a predation method. Still, this method also requires an analytical foundation and, for better communication in the network, also requires a set of mathematical rules. Improving congestion control with scalability, performance, fairness, and stability of the Internet protocol, Analoui and Jamali \cite{98} provided a further extension of their own work \cite{97}. This scheme takes robustness and stability characteristics of biological systems and provides a mathematical model to analyze the performance of the proposed scheme.

\par Analoui and Jamali solved the Internet congestion issue using two nature-inspired techniques \cite{99}. Their paper redefined the problem of congestion control after using the amount of population control method, for effectiveness in the results, they utilized a tri-trophic food chain mathematical model for control. In order to resolve congestion issue, it utilized a queue size of around 10 Kb and achieved 50\% aggregated throughput. On the other hand, the work requires a method to analyze the performance of the system, and needs to consider fairness and stability. In addition, to guarantee achievement in performance and stability, the improved version of bio-inspired congestion control was defined in bio-inspired congestion control-tri-trophic (BICCTT) \cite{100}. Consequently, to design TCP that works as a scalable congestion control with robust and self-adaptive behavior, Hasegawa and Murata \cite{101} presented TCP symbiosis. This work uses the available bandwidth and the physical information about the end-to-end-path to change the TCP window size. Therefore, it utilizes 1000 Mbps bandwidth in 150 sec convergence time.

\par TCP symbiosis tries to resolve the packet loss and unfairness issue using information about bandwidth. It uses two biological models, the Lotka-Volterra model, and the logistic growth model, for adjustment of congestion window rate. The dependencies of these models are equal in changing population sizes that work in living environments. TCP symbiosis uses the window in proportion to the population of a species, and the available bandwidth of a congested link is used as a living environment. A further Linux-based implementation of TCP symbiosis for congestion control was discussed \cite{102}. All the above-defined methods are applicable in Internet environments, and most of the work is related to the adjustment of the data rate after using a mathematical scheme from biological systems. However, these defined schemes are not suitable for hop-by-hop decentralized systems. 

\par To solve congestion-related issues in a network, some methods are defined based on the motivations of insects. Insect methods adopt the shortest pathway from their home to food, which can be used to overcome the congestion-related problems in the network. There are two types of incentive used by insect colonies for tracking food (\textit{the final destination}). First, a chemical substance released by insects is a stimulus for organisms of the same species. Second, the odor of some specific food is also a stimulus to help insects to reach the final destination. If there is no chemical substance released by insects, then the other insects from the same colony will pursue haphazard pathways. The very first stimulus-dependent algorithm was ant-based control (ABC) \cite{103}, and this algorithm was also applied in circuit-switched telephone networks. Ant colony optimization (ACO)-based algorithms work on a traffic awareness rule, which is of great value, as defined in Figure~\ref{10}. ABC algorithms and ACO-based direction-finding work on the same principle. The objectives of the ABC model as the food source and the direction-finding tables are known as pheromone tables. 

\par To improve the quality of calls, ants roaming on crowded paths, will not reach it at an exact time, and ants roaming on fewer congested paths will reach it on time. The sum of impediments is directly proportional to the number of calls used at the same time. In short, calls will not adopt any unsystematic way to arrange themselves in different links. ABC has one negative aspect in that the winner will have to hold on all call setup formats, which means if the most suitable path is congested, then no call can be allocated to it unless there is a change in the probability distribution by the ants. ACO-based algorithms for packet-switched networks will discover different tracks for balancing loads in a pheromone table. 

\subsubsection{\textbf{Ad-hoc Environment-Based Congestion}}
The Lotka-Volterra (LV) competition model explains the link between predators and their prey. If we consider different types of predators in a given set up, then the robustness of one type of predator is influenced by other types that depend only upon some inadequate prey. This feature is applied as a solution to congestion in computer networks. The prediction of wireless sensor networks (WSNs) shows that it has a vast variety of present and future submissions consisting of a large amount (hundreds to thousands) of low-end sensor nodes. The outcome of these large numbers is congestion in the network. This will in turn hinder the consistency of network communications.
 
\par Solving congestion-related issues for WSNs, Antoniou and Pitsillides \cite{104} suggested a method of reducing traffic load step by step by using the LV competition model. In this model, there are different types of nodes playing their parts such as relay nodes (RNs), source-relay nodes (SRNs), and source nodes (SNs). First, data packets are created by SNs and SRNs, which are then transmitted to the non-packet-producing SRNs or RNs. The nodes involved in the network are adaptable and flexible in the transmission of data packets. There is competition between different nodes for the storage capacity to reach the nearby destination. Let us consider WSNs an ecosystem, subdivided into different sub-ecosystems, which are composed of sender nodes forwarding data packets (that behave as competing species) to receiver nodes. These data packets behave as competing species, and their storage capacities are considered insufficient. Traffic flow is described as population bulk that is influenced by connections between indulged streams and storage capacity. 

\par The LV model has two basic features: first, all groups live together, and second, only one group remains alive, fighting with others for their survival, and expressing disapproval of extermination. The survival of a species in the system depends upon the suggested theory that proves that when the contest between members of one species is greater than a contest between two different species, the data flow rate of a species aggregates to form the universal and peaceful existence resolution \cite{105}. The contest between members of two species is the same as the connection time, but the expansion of a species is the opposite of connection time, and the data packet flow rate of SRNs and RNs can be observed by using the same method as SNs. There are different advantages to this method in simulations such as high packet release, low uninterrupted delay, scalability, and fairness of conflicting flow rates.

\par Moreover, our suggested plan is better in performance than the TCP-AIMD like rate-based congestion control scheme \cite{106, 107, 108}. In order to gain strength and efficiency in the data packet flow rate for WSNs, these schemes intensify the QoS violated, the long-time converge, and sawtooth rate performance. The constancy of the system is directly proportional to the contest among species, also discussed in this paper. In short, if we want to improve the efficiency and adaptation of our future networks (according to change circumstances) by using the predator-prey stimulated algorithm, then we must select unfriendliness tariffs.

\par For maintenance of a vibrant equilibrium among different parts of a network system having a constant number of users, the predator-prey association theory and biological symbiosis can be useful. Networks having different components that rely on a small number of processions use the predator-prey relationship scheme, and the different components of networks that depend on each other use biological symbiosis (relationships in which two organisms live with each other and depend upon each other). Networks that use this bio-inspired algorithm (BIA), must apply biological schemes. The properties of individuals and the associations between groups of a biological system, respectively, should be applied in the components and associations between the network's systems. For example, WSNs follow a predator-prey BIA to avoid congestion; and peer-to-peer (P2P) networks depend upon the symbiosis BIA for a source request-response management process. 

\par Avoiding realistic congestion using the flock-of-birds model for wireless sensor networks, Antoniou et al. \cite{109} presented an approach called flock-based congestion control (Flock-CC). This scheme controls congestion by adapting the birds' flock orientated movements and the obstacle-avoidance behavior. It avoids congestion and balances the load after utilizing the alternative paths in wireless sensor networks. This scheme behaves dynamically and achieves adaptation in dynamic network size. In Flock-CC, packets behave like birds; flying space is denoted as the sensor network, and obstacle avoidance illustrates the congested area of the network. 

\par The major inspirations for designing Flock-CC were: 1) closely located individuals, attracting and repelling communications, 2) the migration behavior of birds when they move to the poles, and 3) the narrow field of view in a flock of birds. Flock-CC processes packets after using the above bird-flock movement method, and controls congestion based on it, hop-by-hop. The multiple paths for traffic flow utilize the obstacle avoidance concept from flocks of birds. Antoniou et al. defined the initial stages for development of Flock-CC \cite{109, 110}. Moreover, they provided an extension to, and improvement of, Flock-CC through evaluation results \cite{111, 112}. The enhanced scheme is simple and reduces the implementation complexity level after utilizing two affordable parameters, rather than four. Flock-CC provides efficient retransmission, and avoids collisions with a high packet transmission rate. To achieve the maximum throughput, the proposed method get 93\% packet delivery ratio in hybrid environments. 
  
\par Finally, AntNet \cite{113} and AntHocNet \cite{114} protocols work on the principle of the ant-routing model. These representatives discover the network, and replace stored information (in the form of data packets), just as ants discover their surroundings and use indirect ways of interaction by applying mechanisms of indirect coordination between agents. The practical able routing scheme, offered by AntNet, depends upon the arbitrary resolution of initiating mobile representatives towards their targeted nodes. The ACO algorithm can be used to design the shortest distance between source and target by using random techniques, and brings modifications in path-related, path-plotting accesses in the network. These agents then move back to their habitat (just like ants) after achieving their goals. Therefore, all the routing tables of traveling nodes having recent information about targeted nodes are modified. AntNet has the capability of sustaining the probability for generation of new agents, following current traffic conditions to prevent congestion. The proposed method provides higher and more stable throughput $(10^6 bps)$ in the hybrid scenario. Same as AntNet, AntHocNet also covers routing issue using the hybrid method. After providing, AntHocNet evaluate metric in term of average end-to-end delay, packet overhead and delivery ratio. The purposed system achieves 75\% delivery ratio when the speed of the node is 5 m/s. However, the end-to-end average delay is almost 0.05 sec on the 5 m/s speed. Furthermore, AntHocNet gets 15 routing control overhead at the speed of 10 m/s.

\par All these aforementioned solutions are based on a bio-inspired mechanism that solves congestion issues in the Internet and ad-hoc environments. Some of these works solve congestion issues after adjustment of the data rate in the Internet environment. However, these methods are not applicable in an ASNETs environment. Thus, further solutions are required to solve congestion issues in ASNETs by adopting socially-aware routing schemes.

\begin{figure*}
 \centering  
 \includegraphics[width=1.05\textwidth]{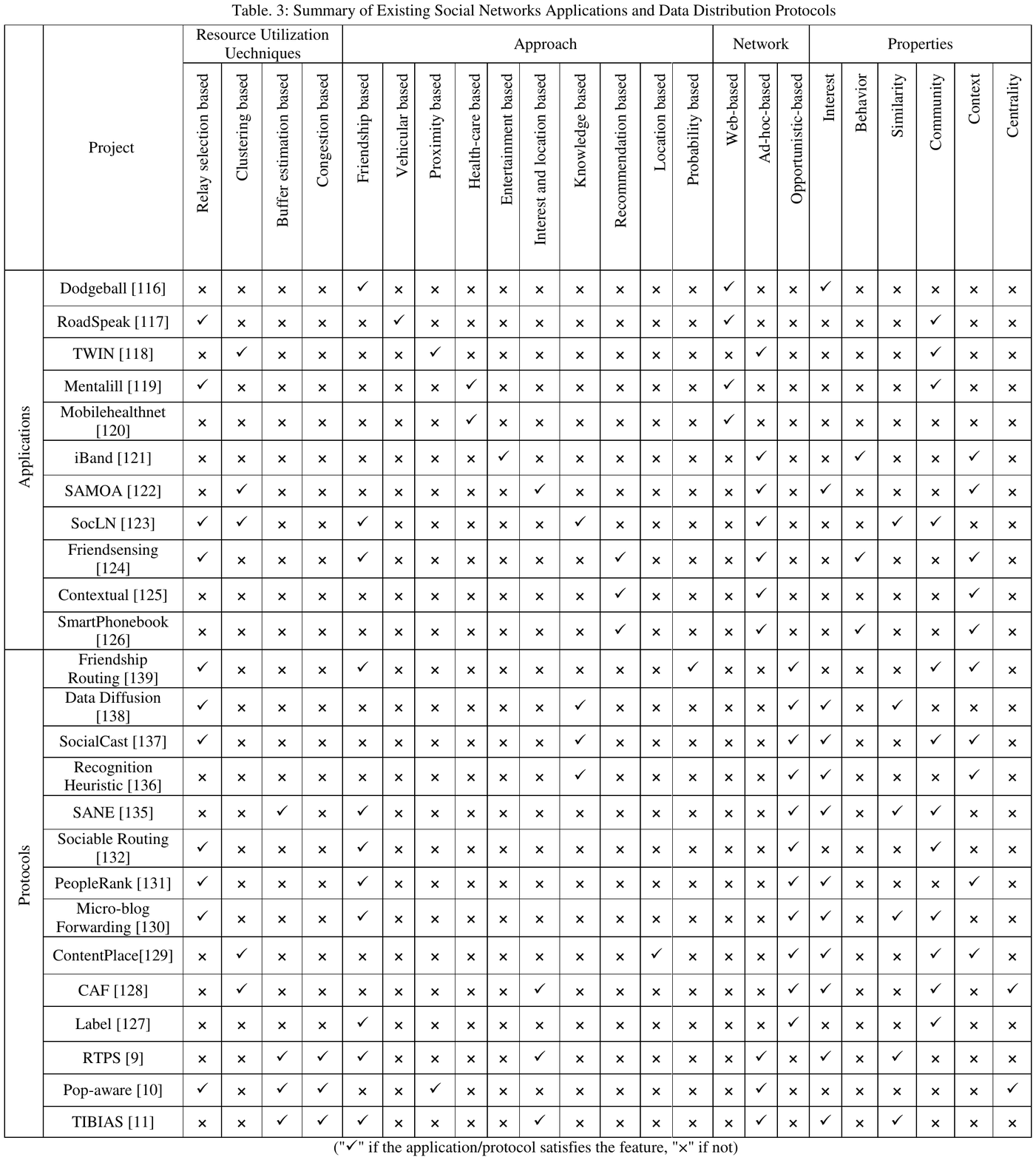}   
 %\caption{Convergence of ad-hoc with social networks.}
% \vspace{-1.5em}
% \label{fig:digraph}
 \end{figure*}

\section{Current Developments In Social Networks}
In this section, we present some state-of-the-art to investigate the developments in social networks, in which nodes use social properties of users to communicate with each other. Mainly, the transfer of data in social networks is designed according to the behavior or interests of the users, in which social properties not only provide the efficient resource utilization but also helpful in determining the mobility patterns which further assist in reducing the congestion-related losses. In order to address the congestion-related losses the authors in \cite{xia2016overhead,xia2016user,liaqat2015social} introduced social awareness-based congestion window adjustment and scheduling techniques. But, these methods are not sufficient enough to resolve congestion issues optimally in ASNETs. Therefore, significant efforts are required to further enhance the existing protocols according to the nature and social properties of the ASNET. 

\par Due to lack of congestion control schemes in ASNETs, we searched for literature on the social network which further helps to design congestion-aware protocols for ASNETs. This section divides the existing methods into two categories. The first related to application developments in social networks, while the second focuses on data distribution techniques designed for socially-aware networks. Table 3 summarizes the existing developments in social networks in both categories. For efficient resource utilization, distribution techniques provide better results in management/replacement of data. However, these techniques do not offer efficient results to control the congestion in ASNETs. This is because, the existing congestion control solutions do not consider dropping, scheduling, data-rate, and control packet-adjustment methods for load balancing. Moreover, in ASNETs, considering the link capacity in a shared network scenario and the link losses is more useful in the congestion management.  

\subsection{Socially-Aware Applications}
The deployment of socially-aware networks includes drawbacks; however, the importance of mobile social applications has promptly increased due to their tremendous value, for instance, their easy and cheap availability, regardless of links and the locations of nodes along social relations. A great deal of variety in these social mobile applications is available and increasing impressively in the market, although, they create congestion in the network. Seada and Perkins \cite{115} proposed an improved model of ad-hoc network that uses a basic architecture for social communications. Following, we describe some important developed applications. Social services cover the applications that are purely related to cellular environments, such as Dodgeball, and are extended to web-based social media forums like LinkedIn \cite{116}. These applications allow members to share their personal data and thoughts, views, and information. Mobile socially-aware applications are also called social services, but the meaning of this specific type is limited solely to social networking services. 

\par On the other hand, a vehicular network application presented by Smaldone et al. \cite{117}, called RoadSpeak, allows users to join communities known as voice chat groups (VCGs). Furthermore, access to the system allows users to exchange voice messages through a central RoadSpeak server that functions as a coordinator. Applications related to mobile multimedia sharing and community-based communication were presented by M\"a\"an\"anen-Vainio-Maattila \cite{118}. The proposed scheme is known as TWIN and it uses social proximity-based communications, with and without familiar persons in ad-hoc networks. Socially-aware applications have been broadly extended to provide assistance in medical facilitation as well. A general architecture for health services was proposed by Chang et al. \cite{119} that uses global positioning system-enabled PDAs. The defined health application connects with a centralized server via general packet radio service (GPRS) and indicates both the sickness and the physical location of the patient. 

\par Additionally, Teles et al. \cite{120} proposed an application called Mobilehealthnet, which works only with some other specific applications. Another wearable socially-aware application helps people using wearable devices to collect and share information regarding social interactions. Besides using wearable applications as behavior modeling and entertainment tools can assist users in daily life activities. Kanis et al. \cite{121} offered a wearable, socially-aware bracelet called iBand, which contains an infrared transceiver for communication purposes and transfers information during a handshake process enabling two iBands. To shift the focus of applications from the virtual to the physical social space, Bottazi et al. \cite{122} presented a socially-aware and mobile architecture (SAMOA), which splits off the social management-related aspects from application requirements for personalizing socially dependent applications. It combines a set of common facility applications. The search for users with the same interest in SAMOA depends on two properties: profile and place visibility.
 
\par To provide optimization in a human intelligence methodology, social learning networks are helpful. In work inspired by human intelligence, Huang et al. \cite{123} proposed a novel application for socially-aware networks. The application allows users to search for common interests and matches neighbors for efficient resource utilization. The proposed method uses the vector space model to design the user's choice based on similarities. This technique is useful for formulating communities in a network, such as reader communities. Besides, a novel application called the recommendation system was designed to provide ease in search techniques. It makes recommendations on various topics after tracking the user's behavior and the availability of data in mobile devices. Two frameworks, SensingHappiness and FriendSensing, that try to find contacts and promote them were proposed by Quercia and Capra \cite{124}. The framework uses short-range radio technologies that are based on logging past encounters. To provide the availability of personalized content to mobile users, Jung \cite{125} put at users' fingertips a social network ontology to detect social relationships. Work related to recommendations to users for call receivers was presented by Min and Cho \cite{126}. However, the communication is based on Bayesian networks. The existing socially-aware applications provide solutions in web and ad-hoc environment. In order to utilize resources efficiently, some of the approaches try to cover it up. However, to minimize the congestion-related loss, none of the above application gives proper solution.

\subsection{Data Distribution Approaches}
The main function and characteristic of opportunistic networks is ``store-carry-forward'', which carries messages between disconnected nodes in the network. Node mobility is one of the major reasons for node disconnection. In node disconnection, a relay node may be used to store and carry the messages until the first node gets reconnected. Hence, the selection for relaying and the distribution of decisions should be based on the specific routing method. Recently, social networks data distribution and sharing protocols have been modified to outperform in the selection of forwarding nodes using social attributes. In this section, we comprehensively discuss the well-known existing socially-aware data distribution schemes.

\par One of the famous, socially-aware distribution schemes, Label was first introduced by Pan and Crowcroft \cite{127}, and is used to employ social characteristics in opportunistic routing \cite{151}. In this approach, every node informs other network nodes about its organization or affiliation in the form of a label. Thus, on the basis of the labels of the potential relay and destination nodes, the approach takes distribution decisions for the node that relate to a similar group as the receivers. The label uses intra-contact time distribution, and inter-contact time distribution to show good distribution nodes from one group to corresponding friend groups. However, the major drawback in the labeling scheme occurs due to large distances among multiple communities. In this scenario, the sender node cannot access the destination node that is involved in its member community. In addition, node mobility increases more dynamically in a network. Thus, in such a dynamic environment, the nodes cannot perform as the best forwarder of data. For example, when nodes are distributed on a network-clustering basis, then the source node cannot communicate with the destination node directly or indirectly. Nevertheless, the paths among clusters are ensured by users that form bridges based on weak associate stalemates. 

\par To solve this issue, Mtibaa and Harras \cite{128} divided large-scale network communities into small communities on the basis of user social interests and their geographic locations. The authors proposed a scheme called community aware forwarding (CAF) to provide efficient results in the distribution of data for large-scale networks. Boldrini et al. introduced ContentPlace \cite{129}, an approach similar to CAF, which forwards data by selecting an appropriate relay node based on community location. In ContentPlace, nodes can be explicitly viewed as a member of more than one locality-related community. For utility-driven content forwarding, the objects of content are defined using two utilities in which one is related to its own preferences, and the other is related to the preferences of communities where the node is a member. This scheme joins the push and pull methods for content distribution. A further enhancement to the primitive push method of ContentPlace was suggested by Allen et al. \cite{130} using socially-aware properties. These authors used the similar interests of users, as well as the frequency in which they meet each other. After the collection of social behaviors, users share social information with each other when other users come within communication range. Another fully distributed scheme called PeopleRank \cite{131} also exploits the social-awareness of users under node mobility. PeopleRank assigns a rank to each user based on a social graph and makes message-distribution decisions using a non-decreasing rank rule. The scheme utilizes consistent socially-aware information among users to decide on the distribution of data. Furthermore, PeopleRank applies the same user level for assigning the next hop in data exchange. This is because socially-aware connected nodes proved to be the best for data distribution. 

\par To forward data among nodes, Fabbri and Verdone introduced a novel sociable routing scheme \cite{132}. In this scheme, for efficient data delivery, the authors select a subset of optimal forwarders, and all nodes rely on these forwarders. The scheme collects social information about the nodes for data routing. However, the social information depends on the frequency and types of encounters. Ning and Yoneki \cite{133}, used socially-aware distribution algorithms, such as Label, Bubble Rap, and PeopleRank \cite{66,67,131}, and investigated their behavior under various opportunistic network configurations. Similar efforts were made by Zyba et al. \cite{134} to investigate the potential and roles of non-social and highly moveable device communications and consequently, users are classified based on their social mobility behavior. For user classification, the total user population is divided into social and vagabond groups. The advantage of Vagabonds is that the group have minimum contact durations, compared to social contacts, which describes how they have more opportunities for data distribution. However, the mentioned socially-aware distribution techniques are heavily based on the ability to store a large amount of data at forwarder nodes.
 
\par The first distribution mechanism, known as social aware networking (SANE) \cite{135}, combines the advantages of both stateless and socially aware distribution approaches in public switched network (PSN) routing. In SANE, users who have similar interests try to communicate more often. The SANE procedures get triggered whenever a node (say $y$), enters communication range of another node (say $z$). In such a scenario, both nodes exchange their interest profiles and scan their buffers for messages to relay. However, a data distribution paradigm is required where each node contributes some fraction of its resources to forward messages, and to store them depends on the service for the whole distribution method. Thus, to overcome such issues, a novel approach was proposed \cite{136}, which is based on fast and frugal recognition heuristics for content-centric mobile networks by exploiting information on how human brains get important information with incomplete knowledge. Different from the previous solutions, SocialCast \cite{137} suggested that efficient data forwarding not only relies on mobility and social ties, but also on the interests of destination nodes. Thus, SocialCast determines a utility function based on node-location prediction and changes in the degree of connectivity. SocialCast is based on a publish-subscribe model, in which nodes publish content on the networks, and receiver nodes get these published content based on their interests. Hence, the node with a greater utility for a particular interest than the node carrying the current message will get copies based on the content-matching approach.

\par Zhang et al. \cite{138} studied a data diffusion problem aimed at diffusing data messages to all moving nodes. Thus, any node that has an interest in getting these data can easily receive them either from stranger nodes or corresponding friend nodes. Thus, analysis reveals interesting results; for example, to obtain better performance, when a node meets a friend node it must first diffuse the data to a node having the most common interests, and similarly, it diffuses the data to a stranger node that is the most different from its common interests while it meets with stranger nodes. Bulut and Szymanski introduced a friendship based routing scheme \cite{139}, which each node distributes data to a node that contains the receiver nodes in its friend communities. Friendship is differentiated based on the time of day, and thus, a node may contain different friendship communities at different times of the day. 

\begin{figure*}[t!]
 \centering  
 \includegraphics[width=.65\textwidth]{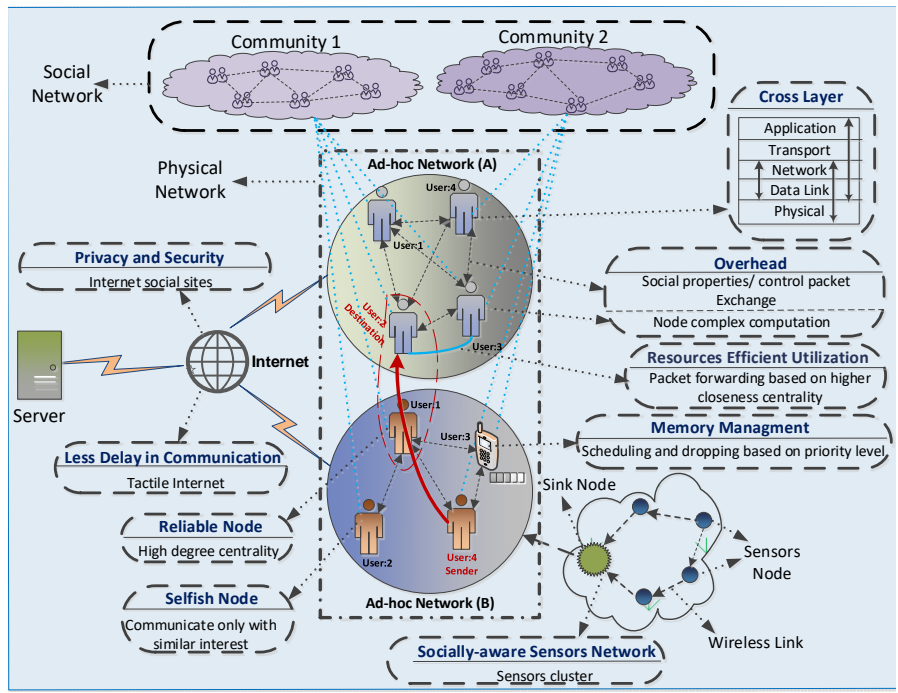}   
 \caption{Research Challenges in ASNETs Environment.}
 \vspace{-1.5em}
 \label{11}
 \end{figure*}
 
\section{Research Challenges And Future Directions}

The above sections provided the detailed working of traditional congestion control protocols under various network environments. However, due to the unavailability of sufficient literature on congestion control methods for ASNETs, the above sections also discussed the existing developments in socially-aware networks such as opportunistic networks where some socially-aware congestion control methods are available. However, these methods do not work accurately in ASNETs environment, where end-to-end connectivity is a prime factor. Moreover, the existing opportunistic network does not consider loss differentiation and share link methods for the adjustment of the congestion window. Hence, our study pointed that research in ASNETs is at its beginning stage specifically in the area of resource efficient and congestion-aware protocols; therefore, significant research efforts are required in designing resource efficient and congestion-aware protocols for ASNETs. Thus, the following sub-sections present some open research challenges in designing resource efficient and congestion-aware protocols for ASNETs. The solutions to these research challenges will be helpful in enhancing the performance of ASNETs and in polishing the node-communication capabilities in scarcely congested environments. Figure 11 maximally cover the graphical description of the presented research issues and their implementation procedures.

\subsection{Overhead Control Mechanisms}
The performance of ASNETs depends on the overhead mechanisms and control of overhead plays a vital role to overcome the congestion issues. In ASNETs, overhead is created in two ways (as shown in Figure 11): when 1) the network communication, and when 2) the node base communication occurs. Firstly, network base overhead occurs when each node exchange social properties (control packets) for communication. However, the cost of overhead increases, when a large number of nodes exchange a huge number of control packets over ASNETs. Therefore, a specific mechanism is required, such as combine acknowledgment to reduce overhead cost on network side communication. Secondly, the complexity/overhead cost of the node communication increases when a large number of nodes share control packet with a single node. Additionally, the high computation cost of an algorithm can also degrade the performance of a node. Therefore, a proper routing mechanism is needed to overcome the problems that may arise from the exchange of a large number of control packets with a single node. Furthermore, in ASNETs, simple algorithms are required that have less complexity while performing efficiently in a short time period. 

\subsection{Selfishness Behavior}
The communication pattern among nodes in ASNETs is different as compared to the traditional ad-hoc networks. Therefore, traditional mechanisms cannot perform better in this environment. In traditional network communication, if the relay node has resources then it will share maximum resources with other nodes. However, in ASNETs, communication parameter affects due to the involvement of social behavior. In order to this as shown in Figure 11, the relay node that has the maximum resources, the relay node may not want to share its resources with other nodes due to selfish behavior. This is because nodes in ASNETs only want to communicate with those nodes that have similar interest or give some advantage in the future. Therefore, to overcome this issue a proper method requires that consider traditional methods with the combination of social schemes (e.g., the use of a credit base or  Tit-for-Tat schemes).

\subsection{Reliable Communication}
For reducing the complexity of the network, social properties help in dividing a larger community into different sub-communities as shown in Figure 11. However, the partition of a large community into sub-communities can affect the reliability of communication. This is because a sender that wants to communicate directly with the destination node cannot do so. In order to provide reliable communication with destination nodes, the communication pattern depends on the intermediate nodes. Therefore, the selection of a reliable node is a major task in ASNETs. In order to resolve such issue, a social property like degree centrality can provide a reliable solution for selection of a relay node. To provide reliable communication in ASNETs, following features, such as responsibility, predictability, faith, and resources are required for selection of the intermediate node.

\subsection{Memory Management}
Maximum users that are involved in ASNETs, use mobile devices for communication. However, the mobile devices that are involved in this network have fewer resources in terms of memory and a large number of nodes transfer application data among each other. Therefore, certain assumptions are required in order to control the memory of the nodes. Overflow in a node buffer can be reduced through the congestion control module. In order to reduce buffer overflow, each node should record its neighbor nodes' capacities. The capacity of a neighbor node provides an advantage by reducing the rate of data on the sender side. Furthermore, the effectiveness of a node buffer can be improved by dropping some of the data from the node buffer and scheduling prioritized data packets first. The reason is that the aim of ASNETs is to provide maximum availability of prioritized data packets. Therefore, dropping-and-scheduling schemes can also enhance the performance of a node buffer by dropping useless data from the node and transferring prioritized packets earlier.

\subsection{Less Delay in Communication}
ASNETs communication depends on the social properties of users (e.g., social ties, similarities, human mobility patterns, centrality, community, and social graphs), which are also helpful in the efficient utilization of resources. In a social networking paradigm (as shown in Figure 11), users want to communicate with specific nodes that have a same social property to reduce congestion in a network. Therefore, to find specific users, the social property called similarity has a lot of importance. The considerations of tie-strengths, human mobility patterns, and social graphs also provide advantages in finding a specific person or user. To facilitate in the quick discovery of a previous communicating partner, the contact history has great significance. Furthermore, ASNETs needs improvement in delay when it communicates with the Internet as shown in Figure 11. The less delay in communication will be helpful to design ultra-high throughput for real-time multimedia service communication through Tactile Internet.
  
\subsection{Cross-layer Architectures}
\par To achieve efficient results and high throughput in ASNETs, some cross-layer techniques provide advantages after combing through multiple layer information, such as the application, transport, network, and physical layers. We know that the social properties of users are helpful in designing optimistic applications in socially-aware networks, where resources are scarce and the environment of the network is congested. As with application layers, we can also apply the social properties of users to different layers and achieve an optimal cross-layer design. Furthermore, in order to improve the performance of ASNETs and achieve the desired services based on application design, it is necessary to propose an efficient middle-ware protocol. This middle-ware protocol will be helpful in providing the lower information to the upper layers, working as a bridge between them. To get rapid decisions in congestion control schemes, a middle-ware protocol will enhance the performance of ASNETs.

\subsection{Socially-aware Communications in WSNs}
\par The communications between nodes in ASNETs are based on the social behaviors of users. Therefore, social properties are required to design the algorithms and protocols. To design WSNs in a socially-aware manner, wireless ad-hoc sensing techniques should sense the data from the environment and apply them to achieve social awareness. Additionally, in order to work in a congested environment, these techniques will be helpful in improving the delivery rates of prioritized node data. To fulfill the above requirements, there is a need to investigate the methods of data collection, processing, and evaluation from WSN nodes. To improve the performance of social property-based applications and communication patterns, further research in WSNs is required.  

\subsection{Privacy and Security}
\par In order to provide privacy and security in socially-aware networks, it is necessary to consider a lot of factors to protect users' personal information against malicious attackers. In socially-aware networks, mobile nodes usually transfer personal and location information about the node through a wireless interface. In addition, when these ad-hoc nodes connect with the Internet and use social networking sites for communications among global users, the privacy and security of users can be affected. This is due to the fact that many social networking sites do not provide effective security for users, and are not trustworthy in protecting users' privacy. 

\par Some surveys are available to discuss the recent privacy and security issues on Internet social networking sites \cite{140}, on MANETs, and in mobile sensing \cite{141}, but existing models are not capable of solving the defined issues \cite{142,143,144,145,146,147,148,149,150}. To solve them, it is necessary to design an application or communications process so that as users communicate among themselves, their personal information will never be disclosed to attackers. Hence, techniques like providing anonymity, hiding location information, and access control and authentication methods must be considered. Along with efficient confidential approaches, there is also a need to design methods for the integrity and availability of data. Complicated confidential methods are not the only reasons to affect the efficiency of data integrity and availability; scarce bandwidth due to congestion also affects the performance of integrity and availability parameters.    

\subsection{Efficient Resource Utilization}
\par In ASNETs, nodes have limited energy levels because of low battery power, and the capacity of the network is limited due to scarce bandwidth in the wireless environment. Besides, users in ASNETs have some socially-aware restrictions and prioritizations. Therefore, to overcome the above-mentioned issues in one solution, it is necessary to design efficient algorithms, protocols, and architectures. The existing work aims to solve the above-defined issues separately. However, none of the existing studies considered social properties of the nodes to allocate resources efficiently. 
\par Additionally, one solution in scarce environments can be provided through a social priority level and assigning resources based on the social level of a node. Providing desired resources among nodes, the efficient allocation of bandwidth provides an advantage. Nevertheless, the calculation of social priority can be based on the network environment and the social interactions among nodes. For example, in congested networks, the environment attempts to forward data packets earlier if they that have higher closeness centrality to the sender. Furthermore, in a big conference hall or at parties where a lot of users meet, it is necessary to the allocate resources of nodes based on a similarity level.

\subsection{Optimization in Data Mining}
\par The users of ASNETs utilize contextual information about nodes for applications and services. Accordingly, an optimal mining approach provides efficient real-time communications and an improvement in the quality of socially-aware services and applications. The communications of users in ASNETs are based on context information about a particular device and the interaction levels among users. Consequently, software for data mining should behave in an optimal way to extract information about the data form users. In addition, local node contextual information can be extended after publishing the same contextual information to online social networks. Publishing the same information on the Internet can also increase the battery lifetime and open up security and privacy issues. Furthermore, this method provides optimized and accurate data for other devices or vehicles.         

\section{Conclusions}

In this paper we have provided background into ad-hoc social networks (ASNETs) and its architecture and protocols. ASNETs can be thought of as the marriage of social networks and ad-hoc networks---the availability of rich contextual social information gives an extra degree of freedom that can be used by network designers to enhance the efficiency of the network while also providing enhanced quality of service. We have provided background into the various ASNET architectures (i.e., centralized, distributed, and hybrid) and social networking concepts (such as the various social properties). We have performed a comprehensive review of the congestion control protocol specifications and solutions in ad-hoc and opportunistic networks. We show that there has been relatively little work focused on embedding social awareness into these congestion control frameworks and that more research into these areas is needed. Finally, we also provided a discussion of existing solutions and suggested areas that are wide open for further research. 

\bibliographystyle{IEEEtran}
\bibliography{LateXSource}

\end{document}